\documentclass[12pt,letterpaper]{article}
\pdfoutput=1
\usepackage{jheppub}
\usepackage{amsfonts}
\usepackage[english]{babel}
\usepackage[utf8]{inputenc}
\usepackage[T1]{fontenc}
\usepackage{slashed}
\usepackage{mathrsfs}
\usepackage{amssymb}
\usepackage{color}
\hypersetup{unicode}
\usepackage{natbib}
\usepackage{mathtools}

\newcommand{\eq}{\begin{equation}}
\newcommand{\feq}{\end{equation}}
\newcommand{\be}{\begin{equation}}
\newcommand{\ee}{\end{equation}}
\newcommand{\eqn}{\begin{eqnarray}}
\newcommand{\feqn}{\end{eqnarray}}
\DeclareMathOperator{\sech}{sech}
\DeclareMathOperator\arctanh{arctanh}

\title{Exact Spherical Branes and Deformations: BMN, IKKT and $\mathcal{N}=2^\ast$}

\author{Cameron Gibson}

\affiliation{George P. \& Cynthia Woods Mitchell Institute for Fundamental Physics and Astronomy,
Texas A{\&}M University, College Station, TX 77843, USA.}

\emailAdd{camerongibson@tamu.edu}
\abstract{We present a unified analytic solution to the BPS equations of spherical Dp-branes for $1\leq p \leq 5$ and give the corresponding type II uplifts in closed form. These are holographically dual to maximally supersymmetric Yang-Mills theory on $S^{p+1}$. The change of variables that makes the BPS equations integrable also allows us to analytically calculate a universal result for the holographic Wilson loop and obtain UV asymptotic behaviour previously determined numerically. The same change of variables extends naturally, with suitable modifications, to the cases $p=0$ and $p=-1$. We obtain a 1-parameter family of D0-brane solutions and show how the IR conditions lead to a singular limit of the integration constant. We also solve the D(-1)-brane BPS equations and show the resulting Euclidean uplift fails regularity conditions. Separately, we obtain an explicit representation of the one-ball electrostatic solution of polarised IKKT, which lets us express multipole moments and holographic observables using elliptic invariants. Finally, we analytically solve the BPS equations of the holographic dual of $\mathcal{N}=2^\ast$ supersymmetric Yang-Mills theory on $S^4$, present the corresponding type IIB uplift and calculate the holographic Wilson loop.}
\begin{document}
\maketitle

\section{Introduction}
The AdS/CFT correspondence was originally proposed by Maldacena \cite{Maldacena:1997re} between the AdS$_5 \times S^5$ near-horizon geometry of N coincident D3-branes and 4D $\mathcal{N}=4$ supersymmetric Yang-Mills (SYM) theory. An extension to $Dp$-branes for $p\neq 3$ was then proposed \cite{Itzhaki:1998dd}. Since then, subsequent work has developed a nonconformal holographic dictionary \cite{Kanitscheider:2008kd,Kanitscheider:2009as,Wiseman:2008qa}. Foundational work by Blau \cite{Blau:2000xg} constructed the Lagrangian of SYM on curved spaces of dimension $d\leq 7$ admitting Killing spinors. Building on previous work by Pestun on supersymmetric localisation on $S^4$ \cite{Pestun:2007rz}, Minahan and Zabzine then gave a unified approach to localisation of maximal SYM on $S^{p+1}$ for $1 \leq p \leq 6$ \cite{Minahan:2015jta}.

In 2018, Bobev, Bomans and Gautason constructed solutions of 10D supergravity which describe the backreaction of Dp-branes with spherical worldvolumes, holographically dual to $(p+1)$-dimensional maximally supersymmetric Yang-Mills (MSYM) theory on $S^{p+1}$ \cite{spherical1}. Since then, free energy and BPS Wilson loop vacuum expectation values (vevs) from field theory have been matched to calculations using the spherical brane supergravity solutions \cite{Bobev:2019bvq} and more recent work has studied $\frac{1}{2}$-BPS Wilson loops beyond leading order \cite{Astesiano:2024sgi}. The BPS equations for these solutions were solved case-by-case and, although analytic solutions in several dimensions have been found, no universal solution has been found.

The first main result of this paper is to provide a universal solution for the spherical brane BPS equations in all dimensions $1 \leq p\leq 5$ using a change of variables. This same change of variables extends naturally to the cases $p=0$ and $p=-1$, which are dual to the BMN and IKKT models, respectively. This gives analytic lower-dimensional solutions across these dimensions and their 10D uplifts. These variables also dramatically simplify holographic calculations such as the holographic Wilson loop. 

The D0-brane case is particularly interesting because of its relation to the BMN model \cite{Berenstein:2002jq,spherical2}. We solve the BPS equations analytically and show the IR boundary condition arises only in a singular limit of the integration constant. We interpret the solution in the electrostatic framework of Lin and Maldacena \cite{Lin:2005nh}. Our solution also provides an analytical description of the BPS scalar trajectories approached near the endpoint of the thermal branch studied by Santos in recent work \cite{Santos:2026rrx}.

We consider the spherical D(-1)-brane system associated with the IKKT matrix model \cite{ikkt,ikktdual1,ikktdual2,ikktdual3}, which was recently included as a special case of a consistent supergravity truncation \cite{henning,henning2}. We analytically solve the BPS equations of \cite{henning} and show that the resulting Euclidean type IIB uplift is not regular using the electrostatic formulation of constructing solutions dual to the polarised IKKT model \cite{komatsu1,komatsu2,polarisedikkt}. Separately, we obtain an explicit representation of the one-ball electrostatic construction of \cite{komatsu1,komatsu2} using a conformal map from the mathematical literature \cite{doubleslit}. This solution falls within the framework of the general solutions by D'Hoker et al. \cite{gutperle2} and uses a conformal-mapping approach also used in recent independent work \cite{Najafian:2026wld}. The benefit of our specific choice of conformal map is that it allows us to write multipole moments and holographic quantities such as the on-shell action explicitly in terms of elliptic invariants and quasi-periods without leaving unevaluated integrals.

We discuss the spherical D5-brane solution and its S-dual NS5-brane solution, which is related to little string theory \cite{lst1}. We derive the IR and UV asymptotics of the NS5-brane solution, including the characteristic linear-dilaton behaviour in the UV. We also show the uplift of the S-dual NS5-brane solution coincides with a limit of the supergravity solutions dual to polarised IKKT found in recent independent work \cite{Najafian:2026wld}.

Our second main result concerns the holographic dual of the non-conformal but still supersymmetric $\mathcal{N}=2^\ast$ SYM on $S^4$. The holographic supergravity solution on $\mathbb{R}^4$ was obtained in \cite{Pilch:2000ue} and Pestun calculated physical observables of $\mathcal{N}=2^\ast$ SYM theory placed on $S^4$ using supersymmetric localisation \cite{Pestun:2007rz}. The holographic dual of the theory on $S^4$ was then constructed in \cite{twostar1}, with an uplift to a type IIB supergravity solution constructed in \cite{Bobev:2018hbq}. The corresponding BPS equations were previously studied numerically and perturbatively (see also \cite{Kim:2019feb}). In this paper, we solve the BPS equations analytically, which allows us to write the uplifted solution analytically, and directly calculate the holographic Wilson loop. We also discuss how applying a similar technique to the holographic dual of equal-mass $\mathcal{N}=1^\ast$ on $S^4$ \cite{Bobev:2016nua} seems to simplify the BPS equations, although we do not obtain a general solution.

This paper is organised as follows. Sections \ref{sec:one}-\ref{sec:three} briefly review the spherical brane construction, derive the universal analytic solution to the BPS equations and present the uplifted solutions, then calculate holographic Wilson loops and UV asymptotic behaviour. Section \ref{sec:five} discusses the D0-brane solution, its electrostatic interpretation, and its relation to non-supersymmetric deformations. Section \ref{sec:four} discusses the D(-1)-brane solution and its electrostatic interpretation, then presents an explicit representation of the one-ball solution of Komatsu et al., with App.~\ref{app:oneball} giving a detailed derivation. Section \ref{sec:six} discusses the D5-brane solution, its S-dual NS5-brane solution and their asymptotic behaviour. Section \ref{sec:nine} presents the analytic $\mathcal{N}=2^\ast$ solution, its IIB uplift and an evaluation of the holographic Wilson loop, then discusses applying a similar method to $\mathcal{N}=1^\ast$ systems, with further details in App.~\ref{app:onestar}. Appendix \ref{subsec:3dembedding} provides an embedding of the bosonic D1-brane theory into 3D gauged supergravity.

\section{Spherical Branes}
\label{sec:one}

The spherical brane solutions constructed in \cite{spherical1} are spherical domain-wall solutions of lower-dimensional maximally supersymmetric gauged supergravities. Since these theories are consistent truncations of type II supergravity on $S^{8-p}$, they have an uplift to 10D Lorentzian type II supergravity solutions, with the time-direction transverse to the Euclidean brane worldvolume (see \cite{Hull:1998vg,Hull:1998fh,Hull:1998ym} for related work on branes with Euclidean worldvolume in type II$^\ast$ string theory).

The solutions of primary interest have a smooth cap in the IR and the radial coordinate combines with the $S^{p+1}$ metric to look locally like $\mathbb{R}^{p+2}$. The name IR is given since the length scale of the sphere corresponds to an IR cut-off in the dual theory. In the UV, the background is the same as the near-horizon limit of flat Dp-brane backgrounds with Euclidean worldvolume. In the dual description, at high energies, the theory should look like SYM on flat space. Imposing these IR and UV conditions chooses the regular solutions out of the 1-parameter families of BPS solutions we later discuss.

In this section, we briefly review the lower-dimensional gauged supergravity action for general dimension. We also find a transformation on the scalar variables, which induces a K\"ahler transformation, that relates the variables adapted to the $p<3$ and $p>3$ cases. This will be useful when giving a general solution for the BPS equations.

\subsection{$(p+2)$-dimensional action}

We will describe Dp-branes with worldvolume $S^d$, where $d=p+1$. The construction here is valid for $1 \leq p \leq 5$. In particular, the $p=6$ case (which we will not consider in this paper) must be treated separately as in \cite{spherical1}. The $p=0$ case is constructed as a limit in \cite{spherical2}. This system was extended to the case of $p=-1$ in \cite{henning}.

We take the spherical domain wall ansatz
\begin{equation}
    ds_{p+2}^2=dr^2+e^{2A(r)}d\Omega_{p+1}^2,
\end{equation}
with radial coordinate $r$, warping function $A$ that only depends on $r$, and $d\Omega_{p+1}^2$ the unit metric on $S^{p+1}$. All scalar fields only depend on $r$.

We consider Euclidean gauged supergravity in $(p+2)$-dimensions with an analytic continuation of the usual axion by $\chi\rightarrow i\chi$. After analytic continuation, $\tau$ parameterises an $SL(2,\mathbb{R})/SO(1,1)$ coset and is written as
\begin{equation}
    \tau=i(\chi+e^{-\beta}), \quad \tilde{\tau}=i(\chi-e^{-\beta}).
\end{equation}
Here, $\tilde{\tau}$ is the analytic continuation of the complex conjugate of $\tau$ and not in general still the complex conjugate of the analytically continued $\tau$.

The Euclidean Lagrangian is
\begin{equation}
    \mathcal{L}=\frac{1}{2\kappa_{p+2}^2}\left(R+\frac{3p}{2(p-6)}|d \eta|^2-2\mathcal{K}_{\tau \tilde{\tau}}\partial \tau \cdot \partial \tilde{\tau}-V\right),
\end{equation}
for the K\"ahler potential
\begin{equation}
    \mathcal{ K}=-\log\left(\frac{\tau-\tilde{\tau}}{2i}\right),
\end{equation}
and the scalar potential
\begin{equation}
    V=\frac{1}{2}e^{\mathcal{K}}\left(\frac{6-p}{3p}(\partial_\eta \mathcal{W})(\partial_\eta \widetilde{\mathcal{W}})+\frac{1}{4}\mathcal{K}^{\tau \tilde{\tau}}D_\tau\mathcal{W} D_{\tilde{\tau}}\widetilde{\mathcal{W}}-\frac{p+1}{2p}\mathcal{W}\widetilde{\mathcal{W}}\right),
\end{equation}
where 
\begin{equation}
    D_a=\partial_a+\partial _a \mathcal{K},
\end{equation}
is the K\"ahler covariant derivative.

The superpotentials are given by
\begin{equation}
    \begin{cases}\mathcal{W}_p={-g\left(3\tau e^{\frac{1}{2}\eta}+(6-p)ie^{\frac{3(2-p)}{2(6-p)}\eta}\right) \quad \text{for} \enspace p<3,}\\{\widehat{\mathcal{W}}_p=-g\left(3ie^{\frac{1}{2}\eta}+(6-p)\tau e^{\frac{3(2-p)}{2(6-p)}\eta}\right) \quad \text{for} \enspace p>3,}
    \end{cases}
\end{equation}
and
\begin{equation}
    \begin{cases}\widetilde{\mathcal{W}}_p={-g\left(3\tilde{\tau} e^{\frac{1}{2}\eta}-(6-p)ie^{\frac{3(2-p)}{2(6-p)}\eta}\right) \quad \text{for} \enspace p<3,}\\{\widehat{\widetilde{\mathcal{W}}}_p=-g\left(-3ie^{\frac{1}{2}\eta}+(6-p)\tilde{\tau} e^{\frac{3(2-p)}{2(6-p)}\eta}\right) \quad \text{for} \enspace p>3,}
    \end{cases}
\end{equation}
where $g$ is the $SO(9-p)$ gauge coupling constant.

The scalar fields $X$ and $Y$ are defined as
\begin{equation}
    \begin{cases}
        \tau=ie^{-\frac{p}{6-p}\eta}(X+Y), \quad \tilde{\tau}=-ie^{-\frac{p}{6-p}\eta}(X-Y), \quad \text{for} \enspace p<3\\
        \tau=ie^{\frac{p}{6-p}\eta}(X+Y), \quad \tilde{\tau}=-ie^{\frac{p}{6-p}\eta}(X-Y), \quad \text{for} \enspace p>3.
    \end{cases}
\end{equation}

Note that the conformal case $p=3$ is recovered separately as the limit of the solutions we consider, for which the scalar fields become constant.

\subsection{Map between $p<3$ and $p>3$}
\label{subsec:kahler}
The scalar parameterisations for $p<3$ and $p>3$ are related by the transformation
\begin{equation}
    \tau \rightarrow -\frac{1}{\tau}, \quad \tilde{\tau} \rightarrow -\frac{1}{\tilde{\tau}}. \label{taukahler}
\end{equation}
This induces a K\"ahler transformation, meaning it is of the form $\mathcal{K}\rightarrow \mathcal{K}+\mathcal{J}+\tilde{\mathcal{J}}, \mathcal{W}\rightarrow e^{-\mathcal{J}}\mathcal{W}$. In this case, it is (noting that multiplying $\mathcal{W}$ by $i$ and $\widetilde{W}$ by $-i$ leaves the action invariant)
\begin{equation}
\begin{split}
    &\mathcal{K}\rightarrow \mathcal{K}+\log \tau + \log \tilde{\tau},\\
    &\mathcal{W}_p \rightarrow \frac{i}{\tau}\widehat{\mathcal{W}}_p, \quad \widehat{W}_p\rightarrow \frac{i}{\tau}\mathcal{W}_p.
    \end{split}
\end{equation}

The transformation in Eq.~\ref{taukahler} is equivalent to
\begin{equation}
    X\rightarrow \widehat{X}=\frac{X}{X^2-Y^2}, \quad Y\rightarrow \widehat{Y}=-\frac{Y}{X^2-Y^2}. \label{xykahler}
\end{equation}

Therefore, when solving the $p>3$ BPS equations, we will solve it using the variables $(X,Y)$, then transform to the physical variables $(\widehat{X},\widehat{Y})$ using Eq.~\ref{xykahler}.

\subsection{Consistent truncations}

When constructing the spherical brane solutions in \cite{spherical1}, they started from a maximally supersymmetric gauged supergravity in $(p+2)$-dimensions and performed a consistent truncation preserving an $SO(3)\times SO(6-p)$ subgroup of the $SO(9-p)$ gauge group, then analytically continued the theory to Euclidean signature with a noncompact $SO(1,2)\times SO(6-p)$ gauge group. 

In that work, they commented that they were not aware of an appropriate 3D supergravity theory obtained by a consistent truncation of type IIB supergravity on $S^7$, which would give the spherical D1-brane system. However, they formally continued their equations to $p=1$ and found that solving the lower-dimensional equations gave solutions to the uplifted solutions. In App.~\ref{subsec:3dembedding}, we describe how the bosonic scalar Lagrangian arises as a consistent truncation of maximal 3D $SO(8)\ltimes T^{28}$ supergravity \cite{Fischbacher:2003yw,3dconsistent}. We do not derive the supersymmetry transformations.

\subsection{Boundary conditions}

IR boundary conditions are imposed by requiring that the scalar fields tend to a constant, with these values given by the critical points of the superpotential $\mathcal{W}$ (or $\widetilde{\mathcal{W}}$), with the metric capping off smoothly. UV boundary conditions are imposed by requiring that, for large values of $r$, the solution should asymptotically approach the Euclidean continuation of the flat domain-wall solutions of the type described in, for example, \cite{Boonstra_1999,Cvetic_1999}.
These boundary conditions are
\begin{equation}
    (X,Y)_\text{IR}=\left(\frac{p}{3},\frac{2(3-p)}{3}\right), \quad (X,Y)_\text{UV}=(1,0),
\end{equation}
which map using Eq.~\ref{xykahler} to the physical variables for $p>3$ as
\begin{equation}
    (\widehat{X},\widehat{Y})_\text{IR}=\left(\frac{-p}{(p-2)(p-6)},\frac{2(3-p)}{(p-2)(p-6)}\right), \quad (\widehat{X},\widehat{Y})_\text{UV}=(1,0).
\end{equation}
\section{Exact Spherical Branes in All Dimensions}
\label{sec:two}

Here, we derive the general solution to the BPS equations for $1 \leq p \leq 5$. A change of variables allows us to turn the general BPS equations into an integrable first order equation which admits a solution involving the incomplete beta function. For each integer value of $p$, we get a solution in terms of elementary functions. Although the BPS equations generally admit a 1-parameter family of solutions, we explain that the IR boundary conditions choose a specific member of this family. We then present the uplifted type II supergravity solutions in closed-form. With suitable modifications, the same approach works for the cases $p=0$ and $p=-1$, which we discuss separately in sections \ref{sec:five} and \ref{sec:four}, respectively.

\subsection{General Solution}
Due to the transformation in section \ref{subsec:kahler}, we can solve the BPS equations using the variables adapted for $p<3$ for all $p$, then make the necessary transformation when $p>3$.
The BPS equations, as in \cite{spherical1}, are
\begin{equation}
    \begin{split}
        &\frac{dY}{dX}=-\frac{Y}{2X}\frac{(p-2)^2+4(1-p)X+7X^2-Y^2}{2(1-X)(2-p+X)-Y^2},\\
        &\frac{d\eta}{dX}=-\frac{6-p}{2pX}\frac{(2-p+X)^2-Y^2}{2(1-X)(2-p+X)-Y^2}. \label{bpseqs}
    \end{split}
\end{equation}
Then we make the change of variables
\begin{equation}
    t=\frac{Y^2}{((2-p)+X)^2}, \quad 1-S=X(1-t).
\end{equation}
The form of $t$ is motivated by the repeated factors of $(3-d+X)$ appearing in the BPS equations, especially the numerator of $\frac{d \eta}{d X}$. Together with the definition of $S$, the first BPS equation becomes
\begin{equation}
\frac{dt}{dS}=\frac{t((2-p)(1-t)+3(1-S))}{2S(1-S)}.
\end{equation}
Setting $q=\frac{1}{t}$, this becomes
\begin{equation}
    \frac{dq}{dS}+\left(\frac{2-p}{2(1-S)}+\frac{5-p}{2S}\right)q=\frac{2-p}{2S(1-S)}.
\end{equation}
This is a first order equation for $q(S)$ which then implies
\begin{equation}
    \frac{d}{dS}\left( q(1-S)^{\frac{p-2}{2}}S^{\frac{5-p}{2}} - \frac{(p-2)}{2}B\left(1-S,\frac{p-2}{2},\frac{5-p}{2}\right)\right)=0, \label{c1}
\end{equation}
where $B\left(1-S,\frac{p-2}{2},\frac{5-p}{2}\right)$ is the incomplete beta function defined by
\begin{equation}
    B(z,\alpha,\beta)=\int_0^z dx \enspace x^{\alpha-1} (1-x)^{\beta-1}.
\end{equation}

Therefore, we can integrate to get the general solution
\begin{equation}
    t(S)=\frac{\frac{2}{p-2}(1-S)^{\frac{p-2}{2}}S^\frac{5-p}{2}}{\frac{2}{2-p}K+B\left(1-S,\frac{p-2}{2},\frac{5-p}{2}\right)} \label{gensol},
\end{equation}
for some integration constant $K$. Note that $p=1$ is understood through analytic continuation of the incomplete Beta function and $p=2$ is understood by taking the finite limit of this expression.

This solution is equivalent to
\begin{equation}
    t(S)=\frac{S^\frac{5-p}{2}}{-K(1-S)^\frac{2-p}{2}+{_2F_1}\left(\frac{p-2}{2},\frac{p-3}{2};\frac{p}{2};1-S\right)},
\end{equation}
which makes manifest the $p=2$ regular limit.

The IR boundary conditions are given by $(t,S)=(1,1)$, which uniquely determines $K=0$. The UV corresponds to $(t,S)=(0,0)$. So we consider the range $0\leq S \leq 1$, with the regular $K=0$ solution being
\begin{equation}
    t(S)=\frac{S^\frac{5-p}{2}}{{_2F_1}\left(\frac{p-2}{2},\frac{p-3}{2};\frac{p}{2};1-S\right)}.
\end{equation}

The original fields $X$ and $Y$ can be written in terms of $S$ as
\begin{equation}
    X(S)=\frac{1-S}{1-t(S)}, \quad Y^2(S)=t(S)(2-p+X(S))^2. \label{xofs}
\end{equation}

From the second BPS equation, we can get the explicit solution for the scalar field $\eta$ as well as its alternative parameterisation by
\begin{equation}
    \rho=e^{-\frac{p}{2}\eta}.
\end{equation}
These can be written in terms of $S$ as
\begin{equation}
    \begin{split}
        &\eta=\eta_0+\frac{(6-p)}{2p}\log\left(\frac{t}{S}\right),\\
          &\rho=\rho_0\left(\frac{t}{S}\right)^\frac{p-6}{4},
    \end{split}
\end{equation}
where we denote the IR values of $\eta$ or $\rho$ as $\eta_0$ or $\rho_0$, respectively.

In our coordinates, we get
\begin{equation}
    ds_{p+2}^2=l_d^2\left(\frac{t}{S}\right)^\frac{p-3}{p}\left(\frac{dS^2}{4S^2(1-S)}+\frac{1-S}{t(S)}d\Omega_d^2\right),
\end{equation}
where
\begin{equation}
l_d=\frac{1}{g}e^{\frac{p-3}{6-p}\eta_0}.
\end{equation}

For the regular solution with $K=0$, the $(p+2)$-dimensional metric caps off smoothly in the IR at $S=1$, where it approaches $\mathbb{R}^{p+2}$.

A curious fact is that if we define $t_p$ as the function $t_p(S)$ for each $p$, there is a recursive relation between $p$ and $(p+2)$. Here, this links $p=1$ and $p=3$, as well as $p=2$ and $p=4$, with $p=3$ and $p=5$ not falling into this relation. It would be interesting to explore whether there is a deeper reason behind this correspondence.
\begin{equation}
    t_{p+2}=\frac{(3-p)(1-S)t_{p}}{p(S-t_{p})}
\end{equation}

We note that our general solution recovers the following previously obtained results after mapping between variables \cite{spherical1}: 
\begin{equation}
\begin{split}
    &t_1(S)=\frac{S^2}{2-S},\\
    &t_2(S)=S^\frac{3}{2},\\
    &t_3(S)=S,\\
    &t_4(S)=\frac{\sqrt{S}+S}{2},
    \end{split}
\end{equation}
with the $p=3$ case understood as a limit.

Our general solution also includes $p=5$ and extrapolates to $p=0$ and $p=-1$, which will be discussed later.

\subsection{Uplift to Type II}

Here, we present the Lorentzian type II solution in 10D that connects the UV and IR in a unified form, valid for $p=1,2,3,4,5$ (with $p=3$ the conformal limit). We use the uplift constructed in \cite{spherical1}, following the convention of having purely imaginary R-R-fields with the usual kinetic-term signs.

The 10D metric in string frame is
\begin{equation}
    ds_{10}^2=\frac{e^{\frac{p}{6-p}\eta_0}}{g^2}\sqrt{\frac{t}{QS}}\left(\frac{dS^2}{4S^2(1-S)}+\frac{1-S}{t}d\Omega_d^2+d\theta^2+P\cos^2\theta d \widetilde{\Omega}_2^2+Q\sin^2 \theta d \Omega^2_{(5-p)}\right),
\end{equation}
where $d \widetilde{\Omega}_2^2$ is the metric of 2D de-Sitter space with volume form $\widetilde{\text{vol}}_2$ and $d\Omega_n^2$ is the round metric on $S^{n}$ with volume form $\text{vol}_{n}$. As explained in \cite{spherical1}, for $p=1$ and $p=2$, regularity requires a double analytical continuation $\theta\rightarrow \frac{\pi}{2}+i \tilde{\theta}$ and $\psi \rightarrow i \tilde{\psi}$ where $\psi$ is a coordinate on dS$_2$. This moves the time coordinate from the de Sitter factor to $\theta$.

The solution can be written explicitly in terms of the coordinates $S$ and $\theta$.
\begin{equation}
\begin{split}
&t(S)=\frac{S^\frac{5-p}{2}}{{_2F_1}\left(\frac{p-2}{2},\frac{p-3}{2};\frac{p}{2};1-S\right)},\\
    &P=\frac{1-S}{(1-S)\sin^2\theta+((1-S+t(p-2))^2-(2-p)^2t)\cos^2 \theta},\\
    &Q=\frac{1-S}{1-t\sin^2 \theta-S\cos^2\theta },\\
    &e^{2\Phi}=g_s^2e^{\frac{p(7-p)}{6-p}\eta_0}\left(\frac{t}{S}\right)^{\frac{7-p}{2}}PQ^\frac{1-p}{2}.
    \end{split}
\end{equation}

We also have form-fields
\begin{equation}
\begin{split}
&B_2=e^{\frac{p}{6-p}\eta_0}\frac{t|1-S+(2-p)(1-t)|P}{g^2\sqrt{S}(1-S)}\cos^3 \theta \widetilde{\text{vol}}_2,\\
&C_{(5-p)}=ie^{-\frac{p}{6-p}\eta_0}\sqrt{S}\frac{|1-S+(2-p)(1-t)|Q}{g_sg^{5-p}(1-S)}\sin^{6-p}\theta\text{vol}_{5-p},\\
&C_{(7-p)}=\frac{i}{g_sg^{7-p}}(\omega(\theta)+P\cos \theta\sin^{6-p}\theta)\widetilde{\text{vol}}_2\wedge \text{vol}_{5-p},
\end{split}
\end{equation}
where
\begin{equation}
    \omega(\theta)=\int _0^\theta d x \sin^{5-p}x.
\end{equation}

\section{Holographic calculations in all dimensions}
\label{sec:three}

The spherical brane solutions provide holographic predictions for observables of MSYM on $S^{p+1}$ that can be compared with supersymmetric localisation. Here, we analytically evaluate the vacuum expectation value of the $\frac{1}{2}$-BPS Wilson loop in the fundamental representation of the gauge group wrapping the equator of $S^{p+1}$ by evaluating the regularised on-shell action of a probe string wrapping the equator of $S^{p+1}$ in the spherical brane solutions. Previously, the on-shell probe string action was solved in certain cases and evaluated numerically in others \cite{Bobev:2019bvq}. Remarkably, using our change of variables, the integrand of this on-shell action becomes a very simple total derivative, which allows us to analytically calculate the Wilson loop vev for $1 \leq p \leq 5$.

The UV asymptotics of scalar fields were calculated numerically in order to calculate the holographic free energy \cite{spherical2}. We analytically determine this asymptotic behaviour and recover the same results conjectured from numerics.

\subsection{Holographic Wilson Loop Calculation}

For convenience, define
\begin{equation}
    \tilde{\eta}_0=\left(\frac{p}{6-p}\right)\eta_0.
\end{equation}
As explained in detail in \cite{Bobev:2019bvq}, consider a probe fundamental string extending along $r$ and wrapping the equator of $S^{p+1}$, lying at constant $\theta=0$ and a fixed point of dS$_2$. Then the on-shell action of this probe string becomes
\begin{equation}
    S_\text{string}=\frac{1}{\ell_s^2}\int dre^{\eta+A}.
\end{equation}

We introduce a UV cutoff at $\epsilon \ll 1$. In the variables $S$ and $t$ introduced previously, $S_\text{string}$ simplifies to
\begin{equation}
    S_\text{string}=\frac{e^{\tilde{\eta}_0}}{g^2\ell_s^2}\int_\epsilon^1 \frac{1}{2S^\frac{3}{2}}dS=\frac{e^{\tilde{\eta}_0}}{g^2\ell_s^2}\left[-\frac{1}{\sqrt{S}}\right]^1_\epsilon=\frac{e^{\tilde{\eta}_0}}{g^2\ell_s^2}\left(\frac{1}{\sqrt{\epsilon}}-1\right).
\end{equation}
Remarkably, in these coordinates, all dependence on $t$ cancels in the integrand.

The counterterm is
\begin{equation}
    S_\text{string,ct}=\left.\frac{1}{g\ell_s^2}e^{A+\frac{3}{6-p}\eta}\right|_{r\rightarrow \infty}=\frac{e^{\tilde{\eta}_0}}{g^2\ell_s^2}\left.\sqrt{\frac{1-S}{S}} \right|_{S=\epsilon}\longrightarrow\frac{e^{\tilde{\eta}_0}}{g^2\ell_s^2}\left(\frac{1}{\sqrt{\epsilon}}\right)+\mathcal{O}\left(\epsilon^\frac{1}{2}\right).
\end{equation}

Therefore, as expected, subtracting the counterterm and taking the UV cutoff to zero leads to the renormalised string action
\begin{equation}
    S^\text{Ren.}_\text{string}=-\frac{e^{\tilde{\eta}_0}}{g^2\ell_s^2},
\end{equation}
which gives the Wilson loop expectation value
\begin{equation}
    \log\langle W(C)\rangle=-S^\text{Ren.}_\text{string}=\frac{e^{\tilde{\eta}_0}}{g^2\ell_s^2}
\end{equation}

\subsection{UV asymptotics}
For $1 \leq p\leq 4$, in the UV limit $S\rightarrow 0$,
\begin{equation}
t\sim \frac{\Gamma(\frac{3}{2})}{\Gamma(\frac{p}{2})\Gamma(\frac{5-p}{2})}S^\frac{5-p}{2}.
\end{equation}

In the same limit, for $p=1,2,4$, ($p=3$ is a special case with $Y=0$),
\begin{equation}
Y^2\sim (3-p)^2S^\frac{5-p}{2}\frac{\Gamma(\frac{3}{2})}{\Gamma(\frac{p}{2})\Gamma(\frac{5-p}{2})}.
\end{equation}

For $p=1$ and $p=2$, this then gives
\begin{equation}
Y^2\sim(3-p)^2(1-X)^\frac{5-p}{2}\frac{\Gamma(\frac{3}{2})}{\Gamma(\frac{p}{2})\Gamma(\frac{5-p}{2})}.
\end{equation}
This analytic expression precisely matches the prediction of \cite{spherical2} from numerical results, which enters into their calculation of the holographic free energy.

Note that $p=4$ is a special case with $\widehat{X}=1$ on the entire trajectory. We deal with $p=5$ separately in Section \ref{sec:uvlst}.

\section{D0-Branes: BMN Model}
\label{sec:five}

The spherical D0-brane system was studied in \cite{spherical2} in the context of holographic duality with a Euclidean version of the BMN matrix model. There, the BPS equations were analysed by expanding around the IR and UV and using numerical integration, without finding a general analytic solution. Here, we integrate the $p=0$ BPS equations and get a 1-parameter family of analytic solutions. This allows us to show that the IR boundary conditions arise as a singular limit of the integration constant, which agrees with the findings of \cite{spherical2}. We then put the solution into the electrostatic framework of Lin and Maldacena \cite{Lin:2005nh}. Finally, we put the BPS solution into the variables of \cite{Santos:2026rrx}, which studies non-supersymmetric deformations of this BPS limit. We comment on some limits of our solution relevant to that work.

\subsection{BPS Solution}

We make the same change of variables as before but now for $p=0$, 
\begin{equation}
    t=\frac{Y^2}{(X+2)^2}, \quad 1-S=X(1-t).
\end{equation}
The boundary conditions are
\begin{equation}
(X,Y)_\text{IR}=(0,2), \quad \quad (X,Y)_\text{UV}=(1,0),
\end{equation}
which is equivalent to
\begin{equation}
    (t,S)_\text{IR}=(1,1), \quad \quad (t,S)_\text{UV}=(0,0).
\end{equation}

The general solution for $t(S)$ in Eq.~\ref{gensol} is ill-defined for $p=0$ but the underlying differential equation in Eq.~\ref{c1} is solvable. It integrates to
\begin{equation}
    t=\frac{S^\frac{5}{2}}{(1-S)(K+\sqrt{S}(3-2S)(1-S)^{-1}-3\arctanh{\sqrt{S}})},
\end{equation}
where $K$ is an integration constant.

As explained in \cite{spherical2}, taking the limit $p\rightarrow 0$ requires us to define
\begin{equation}
    \rho=e^{-\frac{p}{2}\eta},
\end{equation}
and conformally rescale the metric as
\begin{equation}
    ds_2^2=\lim_{p\rightarrow 0} \left(\rho^{\frac{7}{9}-\frac{2}{p}} (1-t)^\frac{1}{3}(1-S)^{-\frac{1}{3}}  ds^2_{p+2}\right).
\end{equation}
This gives
\begin{equation}
    \rho=\rho_0 \left(\frac{S}{t}\right)^\frac{3}{2},
\end{equation}
and
\begin{equation}
    ds_2^2=\frac{\rho_0^\frac{4}{9}}{g^2}\left[\frac{S^2(1-S)^2(1-t)}{t^5}\right]^\frac{1}{3}\left(\frac{tdS^2}{4S^2(1-S)^2}+d\tau^2\right).
\end{equation}

For finite $K>0$, the UV limit $S\rightarrow 0$ gives
\begin{equation}
    t=\frac{S^\frac{5}{2}}{K}+\mathcal{O}\left(S^\frac{7}{2}\right), \label{tnearzero}
\end{equation}
which leads to the correct UV boundary condition.

The IR behaviour is more subtle. To consider how the solution approaches the IR, consider
\begin{equation}
    S=1-\epsilon, \quad \epsilon \ll 1.
\end{equation}
Evaluating $t$ at $S=1-\epsilon$ gives
\begin{equation}
    t(1-\epsilon)=1+\epsilon\left(-K-4+\log 8-\frac{3}{2}\log \epsilon\right)+\mathcal{O}(\epsilon^2 \log^2 \epsilon). \label{tnear1}
\end{equation}

Then
\begin{equation}
    X=\frac{1-S}{1-t}=\frac{1}{K+4-\log 8+\frac{3}{2}\log \epsilon}+\mathcal{O}(\epsilon).
\end{equation}

For $K>0$, Eq.~\ref{tnearzero} implies that $t<1$ close to the UV and Eq.~\ref{tnear1} implies that $t>1$ sufficiently close to $S=1$. Therefore, there is some $0<S^\ast<1$ such that $t(S^\ast)=1$ and the 2D metric becomes singular. Therefore, no regular solution with finite $K>0$ can link the IR to UV. 

Since no finite $K$ gives the correct IR behaviour and taking a limit of $K\rightarrow \infty$ for finite $0<S<1$ gives $t=0$, we consider instead the double-scaling limit
\begin{equation}
    K\rightarrow \infty, \quad S=1-\frac{x}{K}, \quad \text{ for } x \text{ constant, finite}. \label{dsl}
\end{equation}
In this limit,
\begin{equation}
    t\rightarrow \frac{1}{1+x}+\mathcal{O}\left(\frac{\log K}{K}\right).
\end{equation}

Back in the variables $(X,Y)$, we get the leading behaviour 
\begin{equation}
    X\sim \frac{1+x}{K}, \quad Y^2 \sim \frac{4}{1+x}.
\end{equation}
The IR boundary condition $(X,Y)=(0,2)$ is correctly obtained at the limit $x \rightarrow 0$.

In the double scaling limit \ref{dsl},

\begin{equation}
    ds_2^2=g^{-2}xK^{-\frac{2}{3}}(1+x)^{\frac{4}{3}}\rho_0^\frac{4}{9} \left(\frac{(1+x)^{-1}dx^2}{4x^2}+d\tau^2\right).
\end{equation}

For $x=s^2$ and, as $s\rightarrow 0^+$, $dx^2=4s^2ds^2$
\begin{equation}
    ds_2^2=g^{-2}K^{-\frac{2}{3}}\rho_0^\frac{4}{9} \left(ds^2+s^2d\tau^2\right),
\end{equation}
which collapses smoothly if $\tau$ has $2\pi$-periodicity and $g^{-2}\rho_0^\frac{4}{9}K^{-\frac{2}{3}}$ is non-zero and finite. Therefore, we have a smooth IR geometry locally resolved for $1-S=\mathcal{O}(K^{-1})$. However, this does not provide a smooth geometry outside this infinitely thin region near $S=1$. Furthermore, for finite $g$, the scalar field $\rho$ diverges at the IR.

These conclusions are consistent with the findings of \cite{spherical2} obtained through numerical integration and IR and UV expansions.
\subsection{Mapping to LM}

In \cite{spherical2}, the spherical D0-brane system was related to the electrostatic framework of Lin and Maldacena \cite{Lin:2005nh}. Here, we use our BPS solution to obtain an analytic expression for the electrostatic potential. In the electrostatic framework, the supergravity fields are determined by a potential satisfying the axisymmetric 3D Laplace equation
\begin{equation}
    \frac{\partial^2 V}{\partial \rho_{\text{LM}}^2}
    +\frac{1}{\rho_{\text{LM}}} \frac{\partial V}{\partial \rho_{\text{LM}}}
    +\frac{\partial^2 V}{\partial \eta_{\text{LM}}^2}=0.
\end{equation}
We introduce the coordinates $(\xi,\nu)$ through 
\begin{equation}
    \eta_\text{LM}=-ic \nu \xi, \quad \rho_\text{LM}=ic\sqrt{\xi^2-1}\sqrt{1-\nu^2},
\end{equation}
where $1<\xi<\infty$ and $-1 < \nu <1$. These are related to our coordinates by
\begin{equation}
    \xi=\frac{1}{\sqrt{S}}, \quad \nu=\cos \theta, \quad c=\frac{1}{4g^2 \rho_0^\frac{1}{3}}.
\end{equation}
Note that the coordinates of Lin and Maldacena have been analytically continued to $\xi $ and $\nu$.

The electrostatic potential $V$ can then be written as
\begin{equation}
    V=\frac{\rho_0^\frac{2}{3}}{30g^3}\left[3\left[P_3(\nu)Q_3(\xi)-P_1(\nu)Q_1(\xi)\right]+K\left[P_1(\nu)P_1(\xi)-P_3(\nu)P_3(\xi)\right]\right],
\end{equation}
where $P_n$ and $Q_n$ are Legendre polynomials and Legendre functions of the second kind, respectively. Each term manifestly satisfies the axisymmetric Laplace equation.

The $K$-dependent piece of $V$, which we denote $V_K$, is
\begin{equation}
\begin{split}
V_K=&\frac{\rho_0^\frac{2}{3}K}{120g^3}\left[-25\xi^3\nu^3+15\xi^3\nu+15\nu^3\xi-5\nu\xi\right]\\
=&-8ig^3\rho_0^\frac{5}{3}K\left(\rho_\text{LM}^2 \eta_\text{LM}-\frac{2}{3}\eta_\text{LM}^3\right)+i\frac{\rho_0K}{3g}\eta_\text{LM}.
\end{split}
\end{equation}
Since the uplifted type IIA metric and fields are invariant under $V\rightarrow V+\text{constant} \cdot \eta_\text{LM}$, we can identify $V_K$ with the background potential in \cite{Lin:2005nh} up to normalisation, which asymptotes to the supergravity solution produced by flat coincident D0-branes.

The $K$-independent piece of $V$, which we denote $V_Q$, is
\begin{equation}
    V_Q=\frac{\rho_0^\frac{2}{3}}{10g^3}\left[P_3(\nu)Q_3(\xi)-P_1(\nu)Q_1(\xi)\right].
\end{equation}

As $\xi\rightarrow \infty$, 
\begin{equation}
    V_Q\rightarrow -\frac{\rho_0^\frac{2}{3}\nu}{30g^3\xi^2}+\mathcal{O}(\xi^{-4}),
\end{equation}
so $V_Q$ leaves the leading asymptotic behaviour of $V$ unchanged.

$V_Q$ only diverges at $\xi=1$. As $\xi\rightarrow 1^+$, 
\begin{equation}
    V_Q\rightarrow \frac{\rho_0^\frac{2}{3}}{8g^3}\nu(1-\nu^2)\log (\xi-1)+\text{finite terms}.
\end{equation}
This corresponds to a line of charge on the focal line (at $\xi=1,-1 \leq \nu \leq 1$) with charge density
\begin{equation}
    \lambda=\frac{\pi\rho_0^\frac{2}{3}}{5g^3}(P_3(\nu)-P_1(\nu)).
\end{equation}

Therefore, the spherical D0-brane system can be viewed in analytically continued Lin-Maldacena coordinates as a deformation of the standard D0-brane background by a line of charge. Interestingly, the parameter $K$ controls only the strength of the background.

\subsection{Relation to thermal BMN solutions}

In recent work by Santos \cite{Santos:2026rrx}, they studied non-supersymmetric deformations of the spherical D0-brane system. They present BPS equations in Eq.6.14. in their variables as (we add subscript $S$ to all variables from \cite{Santos:2026rrx} to avoid confusion with our variables)
\begin{equation}
\begin{split}
    &\frac{dP_S}{dy_S}=\frac{6+10y_S-8y_S^2-8y_S^3-(15+10y_S-4y_S^2)P_S+9P_S^2}{3y_SP_S},\\
    &\eta_S=\frac{1+y_S-2y_S^2-\frac{3}{2}P_S}{2y_S^2},\\
    &\phi_S=\rho_S^\frac{1}{3}\sqrt{\eta_S}. \label{santosbps}
\end{split}
\end{equation}

This is equivalent to our spherical D0-brane BPS equations in our variables under the map
\begin{equation}
\begin{split}
&y_S=e^{3v_S}\rightarrow \frac{1-t}{1-S},\\
&\eta_S=\frac{\phi_S^2}{\rho_S^\frac{2}{3}}\rightarrow -\frac{t(S+2t-3)^2}{4(1-t)^2},\\
&P_S\rightarrow \frac{(S+2t-3)(2t^2+St+2S-5t)}{3(1-S)^2}.
\end{split}
\end{equation}

Therefore, after changing coordinates, the general solution to Eq.~\ref{santosbps} is given by
\begin{equation}
    t(S)=\frac{S^\frac{5}{2}}{(1-S)(K+\sqrt{S}(3-2S)(1-S)^{-1}-3\arctanh{\sqrt{S}})}.
\end{equation}
Since we want to consider $\eta_S>0$, we now consider $t<0$ which has $K<0$.

\cite{Santos:2026rrx} defines a constant $C$ through
\begin{equation}
\eta_S \simeq Cy_S, \quad y_S \gg 1.
\end{equation}

So
\begin{equation}
    C=\underset{y_S\rightarrow \infty}{\lim} \frac{\eta_S}{y_S}=-\underset{y_S\rightarrow \infty}{\lim} \frac{t}{y_S}=\underset{y_S\rightarrow \infty}{\lim} \left(1-S-\frac{1}{y_S}\right)=1-\underset{y_S\rightarrow \infty}{\lim}S.
\end{equation}

Since $t(S)$ has a pole as $y_S\rightarrow \infty$, we can write our integration constant $K$ in terms of $C$ as
\begin{equation}
    K(C)=3\arctanh \sqrt{1-C}-\frac{\sqrt{1-C}(1+2C)}{C}.
\end{equation}

For the solution relevant to \cite{Santos:2026rrx}, $t<0$ and $t$ has a pole at $S=S^\ast$ for some $0<S^\ast<1$.

In \cite{Santos:2026rrx}, they find numerically that close to the end of the thermal branch, the solutions have two regions. The first is when $y_S \ll y_\mathcal{H}$, and the solution approaches the supersymmetric solution with $\eta_S=0$. The second is when $y_S$ is of order $y_\mathcal{H}$, $\eta_S$ is of order one and the solution enters the region of $y_S\gg1$ and $\eta_S \simeq C y_S$. We examine these limits in our variables, using the fact that Santos numerically finds $C \simeq y_\mathcal{H}^{-1}$.

The first limit is
\begin{equation}
    1<y_S\ll \frac{1}{C}, \quad C \rightarrow 0,
\end{equation}
for which
\begin{equation}
    \begin{split}
        &t=-C\frac{(y_S-1)^\frac{5}{2}}{y_S^\frac{3}{2}}+\mathcal{O}(C^2 \log C),\\
        &S=1-C\left(\frac{y_S-1}{y_S}\right)^\frac{5}{2}-\frac{1}{y_S}+\mathcal{O}(C^2\log C),\\
        &\eta_S=C\frac{(1+2y_S)^2(y_S-1)^\frac{5}{2}}{4y_S^\frac{7}{2}}+\mathcal{O}(C^2 \log C).
    \end{split}
\end{equation}
In this limit, $\eta_S\rightarrow 0$, $t \rightarrow 0$, and $S\rightarrow 1-\frac{1}{y_S}$ which is a particular BPS solution, reproducing the behaviour seen numerically in \cite{Santos:2026rrx} (which has the particular BPS solution written in Eq.6.18).

The second limit is
\begin{equation}
    x\equiv Cy_S =\mathcal{O}(1), \quad C \rightarrow 0,
\end{equation}
for which
\begin{equation}
    \begin{split}
        &S=1-C\left(1+\frac{1}{x}\right)+C^2\left(\frac{5}{2x}-\frac{3}{2}\log\left(1+\frac{1}{x}\right)\right)+\mathcal{O}(C^3),\\
        &t=-x+C\left(\frac{5}{2}-\frac{3}{2}x\log\left(1+\frac{1}{x}\right)\right)+\mathcal{O}(C^2),\\
        &\eta_S=x+\frac{3}{2}C\left(x\log\left(1+\frac{1}{x}\right)-1\right)+\mathcal{O}(C^2).
    \end{split}
\end{equation}

We see that $\eta_S \rightarrow Cy_S$ as seen numerically in \cite{Santos:2026rrx}.

Therefore, the BPS solution provides a perturbative description of the two limiting regions described in \cite{Santos:2026rrx}. Note that we do not provide a description of the nontrivial blackening factor that emerges.

\section{D(-1)-Branes: IKKT Model}
\label{sec:four}

Komatsu et al. constructed a class of Euclidean type IIB supergravity solutions with $SO(7)\times SO(3)$ isometry and 16 supersymmetries \cite{komatsu1,komatsu2}, dual to the polarised IKKT model. In their framework, finding regular 10D solutions that are dual to polarized IKKT amounts to finding the 4D electrostatic potential of a configuration of conducting 3D balls and a background potential. 

The spherical brane family was extrapolated to the $p=-1$ case in recent work \cite{henning2,henning} by analytic continuation. In \cite{henning}, they presented the BPS equations and uplift. Here, we apply our general spherical brane solution to this case and show that it provides a 1-parameter family of solutions. However, we show that in the electrostatic formulation of \cite{komatsu1}, the uplift fails the regularity conditions needed for a regular 10D solution, although interesting features do appear.

Motivated by solving the simplest case of \cite{komatsu1,komatsu2} as explicitly as possible, we solve the one-ball problem using elliptic functions then extract holographic quantities in terms of elliptic invariants. This solution falls under the general solution of \cite{gutperle2} and uses a conformal mapping as does recent work \cite{Najafian:2026wld}, but our method has the benefit of extracting relevant quantities such as multipoles up to arbitrary order using elliptic invariants (which can be written as infinite sums) without the need for numerical integration. A detailed derivation is given in App.~\ref{app:oneball}.

\subsection{BPS solution}

Here, we discuss the spherical D(-1)-brane case obtained as an extension of the spherical brane system extrapolated to $p=-1$ \cite{henning2}. Importantly, there must be an analytic continuation $Y^2\rightarrow -Y^2$. Under this analytic continuation, the BPS equations in Eq.~\ref{bpseqs} become
\begin{equation}
    \begin{split}
        &\frac{dY}{dX}=-\frac{Y}{2X}\frac{9+8X+7X^2+Y^2}{2(1-X)(X+3)+Y^2},\\
        &\frac{d\eta}{dX}=\frac{7}{2X}\frac{(X+3)^2+Y^2}{2(1-X)(X+3)+Y^2}. \label{ikktbps}
    \end{split}
\end{equation}

We can then solve these equations in precisely the same way as the general case.

Defining
\begin{equation}
    t=-\frac{Y^2}{(X+3)^2},\quad 1-S=X(1-t), \label{ikktcov}
\end{equation}
reduces these equations to
\begin{equation}
\frac{dt}{dS}=\frac{3t(2-S-t)}{2S(1-S)}.
\end{equation}
As in the general case, we can solve this by defining $q=\frac{1}{t}$ and integrating, which gives the solution
\begin{equation}
    t=-\frac{S^3}{3S^2-12S+8-K(1-S)^\frac{3}{2}}, \label{ikktsol}
\end{equation}
for an arbitrary constant $K$.

Then, we can use $t(S)$ to write everything in terms of $S$ as
\begin{equation}
X(S)=\frac{1-S}{1-t}, \quad Y^2(S)=-t\left(\frac{4-S-3t}{1-t}\right)^2, \quad e^\eta=e^{\eta_0}\left(-\frac{S}{t}\right)^\frac{7}{2}.
\end{equation}

Note that the case of $K=0$ gives the particularly simple solution to Eqs.~\ref{ikktbps} 
\begin{equation}Y^2=(X+3)\left(\frac{\sqrt{12X^2-8X+12}-(X^2+3)}{X}\right).
\end{equation}

As $S \rightarrow 0$, $t\rightarrow 0$ for any value of $K  \neq 8$, which corresponds asymptotically to a $SO(10)$-invariant solution that uplifts to the near-horizon limit of coincident D(-1)-instantons. 

We consider the UV-connected Euclidean branch with $X>0$, real $Y$ and $S\rightarrow 0^+$, for which $K<8$. For finite $K$, the denominator of $t$ in Eq.~\ref{ikktsol} changes sign for some $S=S^\ast$, with $0<S^\ast<1$. The metric has no smooth cap and there is a singular dilaton in the resulting Euclidean uplift of \cite{henning}. Note that we do not analyse the alternative real forms and Lorentzian continuations of this theory in \cite{henning2}.

\subsection{Electrostatic formulation}

Similarly to the Lin-Maldacena construction for the BMN dual, the dual of the polarised IKKT model can be formulated as an electrostatic problem \cite{komatsu1}. In particular, the potential $V(\rho,z)$ is an axially symmetric solution of the 4D Laplace equation
\begin{equation}
    \frac{\partial^2 V}{\partial z^2}+\frac{2}{\rho}\frac{\partial V}{\partial \rho}+\frac{\partial ^2 V}{\partial \rho^2}=0,
\end{equation}
or, equivalently, $\rho V$ satisfies the 2D Laplace equation
\begin{equation}
    \left(\frac{\partial^2}{\partial z^2}+\frac{\partial^2}{\partial \rho^2}\right)(\rho V)=0. \label{2dlap}
\end{equation}
This is equivalent to $\rho V$ locally being the real or imaginary part of a holomorphic function with respect to the complex coordinate
\begin{equation}
    Z=\rho + iz.
\end{equation}

As shown in \cite{henning}, the variables $(X,Y)$ can be related to the electrostatic variables explicitly as
\begin{equation}
    \rho=\frac{\mu e^{-\frac{\eta}{7}}\sqrt{X}\sqrt{(X+3)^2+Y^2}}{8g^2Y}\sin \theta, \quad z=\frac{\mu e^{-\frac{\eta}{7}}(3+X)}{8g^2Y} \cos \theta,
\end{equation}
with potential
\begin{equation}
    V=\frac{\mu^4\cos \theta e^{3\eta/7}Y(2(X+1)\cos (2\theta)+X-1)}{64g^2X^{3/2}\sqrt{(X+3)^2+Y^2}}.
\end{equation}

In our variables $(t,S)$, this relation becomes much simpler with
\begin{equation}
(\rho,z)=\left(R\sqrt{\frac{1-S}{S}}\sin \theta,R\sqrt{\frac{1}{S}}\cos \theta\right), \quad \text{for constant }  R=\frac{\mu}{8g^2}e^{-\frac{\eta_0}{7}}.
\end{equation}
Substituting the general BPS solution Eq.~\ref{ikktsol}, the potential then becomes
\begin{equation}
\begin{split}
    \rho V=&\frac{e^{\frac{2}{7}\eta_0}\mu^5\sin 2\theta}{2^{10}g^4S^2}\Big[4S(S-2)+4\cos 2\theta(S(S-8)+8)\\
    &\left.+K\sqrt{1-S}\left(2(S-2)\cos 2\theta+S\right)\right]
    \end{split}
\end{equation}

Defining $S=\sech^2 u$, this becomes
\begin{equation}
\begin{split}
    \rho V=\frac{\mu^5 e^{\frac{2}{7}\eta_0}}{2^{12}g^4}&\left[8(\cosh 4u \sin 4 \theta - 2 \cosh 2u \sin 2 \theta)\right.\\
    &\left.+K(2\sinh 2u \sin 2 \theta-\sinh 4u \sin 4 \theta)\right].
    \end{split}
\end{equation}

Further defining $w=u-i\theta$, this becomes
\begin{equation}
    \rho V=\frac{\mu^5 e^{\frac{2}{7}\eta_0}}{2^{12}g^4}\text{Im}\left[8(2\sinh 2w-\sinh 4w)+K(\cosh 4w - 2 \cosh 2w)\right].
\end{equation}

The potential can then be written as 
\begin{equation}
    \rho V=\text{Im}(F_Q+F_K),
\end{equation}
where
\begin{equation}
\begin{split}
    &F_Q=-\frac{\mu^5e^{\frac{2}{7}\eta_0}}{2^6g^4}\frac{Z}{R}\left(1+\frac{Z^2}{R^2}\right)^\frac{3}{2},\\
    &F_K=\frac{\mu^5 e^{\frac{2}{7}\eta_0}}{2^{12}g^4}K\left(3+12\frac{Z^2}{R^2}+8\frac{Z^4}{R^4}\right).
\end{split}
\end{equation}
Therefore, $\rho V$ manifestly satisfies the 2D Laplace equation Eq.~\ref{2dlap} as it is the imaginary part of a holomorphic function of $Z$.

Asymptotically, as $Z\rightarrow \infty$,
\begin{equation}
    F \sim \frac{\mu^5e^{\frac{2}{7}\eta_0}}{2^{12}g^4}\left((K-8)\left(8\frac{Z^4}{R^4}+12\frac{Z^2}{R^2}+3\right)+4\frac{R^2}{Z^2}-\frac{3}{2}\frac{R^4}{Z^4}+\frac{3}{4}\frac{R^6}{Z^6}+\mathcal{O}(Z^{-8}R^8)\right)
\end{equation}
Therefore, the terms that govern the asymptotic behaviour of $F$ are proportional to $(K-8)$.

The uplifted D(-1)-brane solution also fails to satisfy the regularity conditions of \cite{komatsu1} that would give a regular 10D solution. For example, as $\rho \rightarrow 0$, for $0<z<R$
\begin{equation}
    \rho V \rightarrow -\frac{\mu^5e^{\frac{2}{7}\eta_0}}{2^6 g^4}\frac{z}{R}\left(1-\frac{z^2}{R^2}\right)^\frac{3}{2} \neq 0,
\end{equation}
independently of $K$.

\subsection{One-ball electrostatic solution}
\label{subsec:oneballsol}

In \cite{komatsu1,komatsu2}, a family of Euclidean IIB supergravity solutions was presented dual to the polarised IKKT model. These solutions are entirely determined by a potential $V(\rho,z)$ satisfying the 4D axially symmetric Laplace equation, which is equivalent to $\rho V$ being the real or imaginary part of a holomorphic function of the complex coordinate $Z=\rho+iz$. 

Here, we consider the simplest solution in that setup, which consists of one 3D conducting ball centered at $z=z_s$, with charge $Q_s$, and radius $R_s$ determined by the charge density vanishing at the edge. $V(\rho,z_s)=V_s$ for constant $V_s$ for $0\leq \rho \leq R_s$. There is a grounded plane at $z=0$. Furthermore, the potential should asymptote to the background potential
\begin{equation}
    V_{bg}=-\eta z \mu^3+\frac{\mu^5}{2^7}(z\rho^2-z^3).
\end{equation}

Here, we present an explicit solution of this one-ball construction using a conformal map from the mathematical literature \cite{doubleslit,Koppenfels1959}. A detailed derivation is given in App.~\ref{app:oneball}.

We consider $Z=\rho+iz$ in
\begin{equation}
   Z\in   G=\mathbb{C}\backslash ([-R_s,R_s]\pm i z_s),
\end{equation}
which removes two slits where the conducting ball and its image (to ensure a grounded plane at $z=0$).

We can then write an explicit conformal map \cite{doubleslit,Koppenfels1959}
\begin{equation}
\begin{split}
&\Pi\rightarrow G: v\rightarrow Z;\\
    &Z(v)=-\frac{z_s}{\pi}(\zeta(v)-\eta_1v),\\
    &\text{where }\Pi=\left\{-\frac{1}{2}<\text{Re}(v)<\frac{1}{2},-\frac{m}{2}<\text{Im}(v)<\frac{m}{2}, \quad v \sim v+1\right\}.
\end{split}
\end{equation}
for $\zeta$ the Weierstrass zeta function with periods $\omega_1=1$ and $\omega_2=im$, and $\eta_1\equiv 2\zeta(1/2)$.
The potential can then be written explicitly as
\begin{equation}
\begin{split}
    &\rho V(v)=\text{Re}(F(v)),\\
    &F(Z(v))=\left(\frac{\eta\mu^3}{2}+\frac{3\mu^5z_s^2\eta_1}{2^8\pi^2}\right)iZ(v)^2-\frac{z_s^2 \mu^5}{2^9\pi^2}iZ_v(v)^2.
\end{split}
\end{equation}

We also get the constraint
\begin{equation}
    \wp(v_c)=-\eta_1, \quad
    R_s=\frac{z_s}{\pi}|\text{Re}[\zeta(v_c)-\eta_1v_c]|,
\end{equation}
for the Weierstrass elliptic function $\wp$ and $V_s$ can be written as
\begin{equation}
    V_s=-\eta\mu^3z_s-\frac{3\mu^5 z_s^3\eta_1}{2^7\pi^2}.
\end{equation}

We get an explicit form for the charge
\begin{equation}
Q_s=\frac{z_s^5\mu^5(20\eta_1^3-\eta_1g_2-g_3)}{2560\pi^4},
\end{equation}
and multipoles up to arbitrary order can be calculated, such as
\begin{equation}
\begin{split}&q_3=\frac{(5040\eta_1^4-336\eta_1^2g_2+5g_2^2-216\eta_1g_3)\mu^5z_s^8}{215040\pi^6},\\ &q_5=\frac{\eta_1(504\eta_1^4-42\eta_1^2g_2+g_2^2-18\eta_1g_3)\mu^5z_s^{10}}{10752\pi^8}.
\end{split}
\end{equation}

This then allows, for example, the on-shell action from the matrix model in ~\cite{komatsu2} to be written explicitly in terms of elliptic invariants, detailed in App.~\ref{subsec:onshell}

By construction, the potential vanishes at $z=0$, $V(\rho,z_s)=V_s$ for constant $V_s$ for $0\leq \rho \leq R_s$, and the potential has the desired background asymptotics. The charge density vanishes at the edge of the ball. We expect the solution here to satisfy the regularity requirements of \cite{komatsu1} which lead to a proof of positivity, but do not make a claim here of global regularity and positivity.

\section{D5-Branes: Little String Theory}
\label{sec:six}
D5-branes are a particularly interesting case due to its S-dual NS5-brane solution being linked to little string theory \cite{lst1}, signalled by the linear asymptotics of the dilaton. Here, we present the analytic solution of spherical D5-branes, the UV and IR asymptotics, and the $(p,q)$-family obtained by $SL(2,\mathbb{R})$ transformations.

Recent independent work \cite{Najafian:2026wld} took a limit to obtain an NS5-brane solution and commented that it may be related to the spherical brane system. Interestingly, this solution precisely coincides with the S-dual of the uplifted spherical D5-system (the map from our variables to theirs is simply $\theta \rightarrow \theta+\frac{\pi}{2}$, $u \rightarrow y$, together with a suitable analytic continuation).
\subsection{BPS solution}
\label{sec:uvlst}
The solution is
\begin{equation}
t=\frac{\frac{2}{3}(1-S)^\frac{3}{2}}{2(\arctanh\sqrt{1-S}-\sqrt{1-S})}
\end{equation}
Then we make the natural substitution 
\begin{equation}
    \tanh^2 u = 1-S  \quad \text{or equivalently} \quad S=\sech^2 u,
\end{equation}
so that
\begin{equation}
    t=\frac{\tanh^3 u}{3(u-\tanh u)}.
\end{equation}

For convenience, define
\begin{equation}
    T(u)=u\coth u -1.
\end{equation}

After the transformation in Eq.~\ref{xykahler}, the solution can be written as
\begin{equation}
    \widehat{X}=\frac{T^2}{u^2-3T}, \quad \widehat{Y}^2=\frac{T}{3u^2}(3\widehat{X}-1)^2.
\end{equation}

Note that the IR values $(\widehat{X},\widehat{Y})=(\frac{5}{3},\frac{4}{3})$ correspond to $u\rightarrow 0$ and UV values $(X,Y)=(1,0)$ correspond to $u\rightarrow \infty$.

Furthermore, we can solve for $\eta$ to get
\begin{equation}
\eta(u)=\eta_0+\frac{1}{10}\ln\left(\frac{\sinh^2(u)}{3T(u)}\right).
\end{equation}
The 7D metric is
\begin{equation}
    ds_7^2=g^{-2}e^{4\eta}\left(du^2+3T d\Omega_6^2\right).
\end{equation}
The uplifted 10D metric is 
\begin{equation}
    ds^2=\frac{e^{5\eta}}{g^2\sqrt{Q}}\left(du^2+3Td\Omega_6^2+d\theta^2+P \cos^2 \theta d \tilde{\Omega}_2^2\right),
\end{equation}
where
\begin{equation}
    P=\frac{\widehat{X}}{\cos^2\theta+\widehat{X}\sin^2 \theta}, \quad Q=\frac{\widehat{X}}{\widehat{X}\cos^2 \theta+(\widehat{X}^2-\widehat{Y}^2)\sin^2 \theta}.
\end{equation}

The UV asymptotics are different from other dimensions and can be written exactly as
\begin{equation}
\begin{split}
    &\widehat{Y}^2=\frac{(u-1)(2u-3)^2}{3(u^2-3u+3)^2}+\frac{2(2u-3)(14u^4-51u^3+81u^2-63u+18)
    }{3(u^2-3u+3)^3}e^{-2u}+\mathcal{O}(e^{-4u}),\\
    &\widehat{X}=\frac{(u-1)^2}{u^2-3u+3}+\frac{2u(u-1)(2u^2-3u+3)}{(u^2-3u+3)^2}e^{-2u}+\mathcal{O}(e^{-4u}),\\
    &\eta=\eta_0-\frac{1}{10}\log(12(u-1))+\frac{1}{5}u-\frac{2u-1}{5(u-1)}e^{-2u}+\mathcal{O}(e^{-4u}).
    \end{split}
\end{equation}

\subsection{(p,q) Branes}

Upon moving to the Einstein frame, the uplifted 10D metric is
\begin{equation}
    ds^2=e^{-\frac{\Phi}{2}}\frac{e^{5\eta}}{g^2\sqrt{Q}}\left(du^2+3Td\Omega_6^2+d\theta^2+P \cos^2 \theta d \tilde{\Omega}_2^2\right).
\end{equation}

Define
\begin{equation}
    \tau_\text{IIB}=C_0+ie^{-\Phi}.
\end{equation}

By acting on the uplifted solution with $SL(2,\mathbb{R})$ transformations, it is possible to get a family of $(p,q)$ fivebranes wrapped on $S^6$.
We get
\begin{equation}
    \tau_{\text{IIB}}\rightarrow \frac{a\tau_\text{IIB}+b}{c\tau_\text{IIB}+d}, \quad \begin{pmatrix}
        C_2\\B_2
    \end{pmatrix}\rightarrow \begin{pmatrix}
        a&b\\c&d
    \end{pmatrix}\begin{pmatrix}
        C_2\\B_2
    \end{pmatrix}.
\end{equation}

In particular, we can transform from the D5-brane solution to the NS5-brane solution by leaving the Einstein frame metric unchanged and transforming $\tau_\text{IIB}\rightarrow -\frac{1}{\tau_\text{IIB}}$ and $(C_2,B_2)\rightarrow (-B_2,C_2)$.

Then, moving to the string frame gives
\begin{equation}
    ds^2=\frac{1}{g^2g_s\sqrt{\widehat{X}}}\left(du^2+3Td\Omega_6^2+d\theta^2+P \cos^2 \theta d \tilde{\Omega}_2^2\right).
\end{equation}

The dilaton is
\begin{equation}
    e^{\Phi_\text{NS}}=\frac{e^{-5\eta}}{g_s }\frac{\sqrt{P}}{\widehat{X}}.
\end{equation}

As $u \rightarrow 0$, 
\begin{equation}
    ds^2\rightarrow \frac{1}{g^2g_s}\sqrt{\frac{3}{5}}\left(du^2+u^2d\Omega_6^2+d\theta^2+\frac{5\cos^2 \theta}{4-\cos 2\theta}d \tilde{\Omega}_2^2\right),
\end{equation}
and
\begin{equation}
    e^{\Phi_\text{NS}}\rightarrow \frac{3e^{-5\eta_0}}{g_s\sqrt{5(4-\cos 2 \theta)}}+\mathcal{O}(u^2).
\end{equation}

As $u \rightarrow \infty$,
\begin{equation}
    ds^2\rightarrow \frac{1}{g^2g_s}\left( du^2+3ud\Omega_6^2+d\theta^2+\cos^2 \theta d \tilde{\Omega}_2^2\right),
\end{equation}
and
\begin{equation}
    \Phi_\text{NS}\rightarrow -u+\frac{1}{2}\log (12u)-5\eta_0-\log g_s+\mathcal{O}(u^{-1}),
\end{equation}
with also exponentially subleading terms.

The linear asymptotics of the dilaton is a sign this theory is linked to little string theory \cite{lst1}.

\section{$\mathcal{N}=2^\ast$ on $S^4$ Solution}
\label{sec:nine}

In the conformal limit $p\rightarrow 3$, the general spherical brane solution becomes the near-horizon geometry of Euclidean D3-branes. Here, we analytically solve the BPS equations of the holographic dual of the $\mathcal{N}=2^\ast$ theory on $S^4$, which is a non-conformal but supersymmetric deformation of the spherical D3-brane system \cite{twostar1}. 

We solve the BPS equations by changing coordinates in a way adapted to the scalar coset that simplifies the Lagrangian and BPS equations. We then identify a conserved quantity on a simple invariant subspace of the BPS equations, then show it is in fact conserved in the full system, which leads to a full solution. We then write the corresponding type IIB solution using the uplift in \cite{Bobev:2018hbq}, choosing coordinates similar to the previous spherical brane solutions. We then analytically evaluate an expression for the holographic Wilson loop.

We also discuss how applying the same method to the equal-mass $\mathcal{N}=1^\ast$ case leads to some simplifications but does not lead to an analytic solution.

\subsection{BPS solution}

Here, we use the same notation as \cite{twostar1} to introduce the Lagrangian and BPS equations then solve them analytically.

The solution is a domain wall with $S^4$ slices with the metric
\begin{equation}
    ds^2=dr^2+e^{2A(r)}d\Omega_4^2,
\end{equation}
where $d\Omega_4^2$ is the unit round metric on $S^4$. There are three scalar fields $z(r),\tilde{z}(r)$ and $\eta(r)$, which only depend on the radial coordinate $r$. $z$ and $\tilde{z}$ are treated as independent fields after analytic continuation.

The Lagrangian is
\begin{equation}
    \mathcal{L}=\left[R_5-\frac{12\partial_\mu\eta \partial^\mu\eta}{\eta^2}-4\frac{\partial_\mu z\partial^\mu \tilde{z}}{(1-z\tilde{z})^2}-V\right],
\end{equation}
where
\begin{equation}
    V=-g^2\left(\frac{1}{\eta^4}+2\eta^2\frac{1+z\tilde{z}}{1-z\tilde{z}}+\frac{\eta^8}{4}\frac{(z-\tilde{z})^2}{(1-z\tilde{z})^2}\right),
\end{equation}
with $g$ the 5D gauge coupling.

The BPS equations are then
\begin{equation}
\begin{split}
    &z'=-3(1-z\tilde{z})(\eta')\frac{2(z+\tilde{z})+\eta^6(z-\tilde{z})
    }{2\eta(1+\tilde{z}^2-(1-\tilde{z}^2)\eta^6)},\\
    &\tilde{z}'=-3(1-z\tilde{z})(\eta')\frac{2(z+\tilde{z})-\eta^6(z-\tilde{z})
    }{2\eta(1+z^2-(1-z^2)\eta^6)},\\
    &(\eta')^2=g^2\frac{(1+\tilde{z}^2-(1-\tilde{z}^2)\eta^6)(1+z^2-(1-z^2)\eta^6)}{36(1-z\tilde{z})^2\eta^2},
\end{split}
\end{equation}
where $'$ denotes differentiating with respect to $r$.

The warping function is determined by
\begin{equation}
    e^{A}=\frac{12(1-z\tilde{z})^2(\eta')}{g^2(z^2-\tilde{z}^2)\eta^3}.
\end{equation}

The scalar fields $z$ and $\tilde{z}$ parameterise the analytically continued $SU(1,1)/U(1)$ coset. We make a natural choice of coordinates adapted to the coset space by choosing embedding coordinates in a hyperboloid
\begin{equation}
X^0=\frac{1+z\tilde{z}}{1-z\tilde{z}}, \quad X^1=\frac{z+\tilde{z}}{1-z\tilde{z}}, \quad X^2=\frac{z-\tilde{z}}{1-z\tilde{z}},
\end{equation}
such that 
\begin{equation}
    {(X^0)}^2-{(X^1)}^2+({X^2)}^2=1. \label{hyp}
\end{equation}
This greatly simplifies the potential to be polynomial in $X^i$. Another simplifying change of variable is
\begin{equation}
    Y=\eta^6.
\end{equation}

Doing these leads to
\begin{equation}
\begin{split}
    &\frac{dX^0}{dY}=\frac{Y-Y^2X^0-Y{(X^0)}^2+Y^2{(X^0)}^3+3Y{(X^1)}^2-2X^0{(X^1)}^2-Y^2X^0{(X^1)}^2}{2Y({(X^1)}^2(1-Y^2)+(1-YX^0)^2)},\\
    &\frac{dX^1}{dY}=X^1\frac{-2-Y^2+2YX^0+Y^2{(X^0)}^2-2{(X^1)}^2-Y^2{(X^1)}^2}{2Y({(X^1)}^2(1-Y^2)+(1-YX^0)^2)},\\
    &\frac{dX^2}{dY}=X^2\frac{-YX^0+Y^2{(X^0)}^2-2{(X^1)}^2-Y^2{(X^1)}^2}{2Y({(X^1)}^2(1-Y^2)+(1-YX^0)^2)}.\\
\end{split}
\end{equation}

Since $\frac{dX^2}{dY}$ is proportional to $X^2$, consider the subspace $X^2=0$. Then ${(X^1)}^2={(X^0)}^2-1$ and we get
\begin{equation}
\begin{split}
    &\frac{dX^0}{dY}=\frac{{(X^0)}^2-1}{Y(Y-X^0)},\\
    &\frac{dX^1}{dY}=\frac{X^0X^1}{Y(Y-X^0)}.
\end{split}
\end{equation}
The first equation is solved by
\begin{equation}
    \frac{1}{Y}=X^0+c\sqrt{{(X^0)}^2-1},
\end{equation}
for a constant $c$.
Using the constraint coming from $X^2=0$, this can be written as
\begin{equation}
    \frac{1-YX^0}{YX^1}=c.
\end{equation}
This is then consistent with the second equation.

Perhaps surprisingly, this constant is actually conserved even when $X^2\neq 0$ and is in fact a valid conserved quantity for the full BPS equations.

Now we can use the conserved quantity to eliminate $X^0$ from Eq.~\ref{hyp} to get
\begin{equation}
    \left(\frac{1}{Y}-cX^1\right)^2-{(X^1)}^2+{(X^2)}^2=1.
\end{equation}

Completing the square in $X^1$ gives
\begin{equation}
    \left((c^2-1)YX^1-c\right)^2+Y^2(c^2-1){(X^2)}^2=1+Y^2(c^2-1),
\end{equation}
We can now choose the natural coordinates
\begin{equation}
    R^2=1+Y^2(c^2-1), \quad U=(c^2-1)YX^1-c, \quad V=Y\sqrt{1-c^2}X^2,
\end{equation}
so that
\begin{equation}
U^2-V^2=R^2.
\end{equation}

Writing the BPS equations in these variables gives
\begin{equation}
\begin{split}
    &\frac{dU}{dR}=\frac{(2cU+2U^2+V^2)}{2(c+U)R},\\
    &\frac{dV}{dR}=\frac{(2c+3U)V}{2(c+U)R}\\
    &\left(\frac{dR}{dr}\right)^2=g^2(U+c)^2\left(\frac{R^2-1}{c^2-1}\right)^\frac{2}{3}.
\end{split}
\end{equation}

Parameterise
\begin{equation}
    U=R\cosh u, \quad V=R\sinh u.
\end{equation}
Here, $u$ is a coordinate such that $u\rightarrow 0$ is the UV and $u\rightarrow \infty$ is the IR.
Then
\begin{equation}
    \frac{du}{dR}=\frac{\sinh u}{2(R\cosh u+c)}.
\end{equation}

This is solvable by
\begin{equation}
    R(u)=-c\cosh u+\left(c'+c\log\coth\frac{u}{2}\right)\sinh^2u
\end{equation}
Now we can go back to get

\begin{equation}
    \begin{split}
    &X^0=\frac{1+cR\cosh u}{\sqrt{1-c^2}\sqrt{1-R^2}},\\
        &X^2=\frac{R\sinh u}{\sqrt{1-R^2}},\\
        &X^1=-\frac{c+R\cosh u}{\sqrt{1-c^2}\sqrt{1-R^2}}.\\
    \end{split}
\end{equation}
This gives the general solution to the BPS equations parameterised by $c$ and $c'$
\begin{equation}
    \begin{split}
    &R(u)=-c\cosh u+\left(c'+c\log\coth\frac{u}{2}\right)\sinh^2u,\\
        &z(u)=\frac{\sqrt{1-c^2}R\sinh u-c-R\cosh u}{(1-c^2)Y+cR\cosh u+1},\\
        &\tilde{z}(u)=\frac{-\sqrt{1-c^2}R\sinh u-c-R\cosh u}{(1-c^2)Y+cR\cosh u+1},\\
        &\eta^6=\sqrt{\frac{1-R^2}{1-c^2}},\\
        &e^{2A}=\frac{4(1-c^2)\eta^4}{g^2\sinh ^2 u}. \label{fieldssol}
    \end{split}
\end{equation}

Note that the  integration constant $c'$ in $R(u)$ is fixed by smoothness of the metric at $u\rightarrow \infty$ to be $c'=0$.

\subsection{Uplift to type IIB}

We write the uplift following the construction of \cite{Bobev:2018hbq}. For $|c|<1$ and $|\eta|<1$, we can choose the (Euclidean) metric, axio-dilaton and four-form to be real with purely imaginary two-forms. More general choices can lead to complex 10D fields (see \cite{Bobev:2018hbq} for a more detailed discussion of the signature). We choose slightly different coordinates such that the metric is structurally similar to the spherical brane solutions constructed in this paper. 
Define
\begin{equation}
\begin{split}
    & H=R \cosh u,\\
   & S=\tanh^2 u,\\
    &t=\frac{S}{1-R^2},
\end{split}
\end{equation}
and
\begin{equation}
\begin{split}
&P=1+cH-(c+H)\cos \omega,\\
&Q_1=P\cos^2 \theta+(1-H^2)\sin^2\theta,\\
&Q_2=P\cos^2\theta+\frac{S}{t}\sin^2\theta.
\end{split}
\end{equation}
Then we can write the 10D metric as
\begin{equation}
    ds^2=\left(\frac{4(Q_1Q_2)^\frac{1}{4}}{\eta^6 g^2 \sqrt{g_s}}\right)\sqrt{\frac{t}{S}}\left(\frac{dS^2}{4S^2(1-S)}+\left(\frac{1-S}{t}\right)d\Omega_4^2+\frac{g^2}{4}\eta^8d\Omega_5^2\right).
\end{equation}
$d\Omega_5^2$ is the deformed metric on $S^5$
\begin{equation}
    d\Omega_5^2=\frac{4\eta^4}{g^2}\left(\frac{1-c^2}{P}d\theta^2+\frac{(1-c^2)}{Q_2}\sin^2\theta d\phi^2+\cos^2 \theta\left(\frac{P}{Q_1}(E_1^2+E_2^2)+\frac{(1-c^2)^2\eta^{12}}{PQ_2}E_3^2\right)\right).
\end{equation}

One forms $E_i$ are defined as deformations of the left-invariant one-forms $\sigma_i$ on $SU(2)$.
\begin{equation}
    \begin{split}
    &\sigma_1=-\frac{1}{2}(\sin \xi_-d \omega-\sin \omega\cos \xi_-d\xi_+),\\
    &\sigma_2=-\frac{1}{2}(\cos \xi_-d\omega+\sin \omega\sin \xi_-d \xi_+),\\
    &\sigma_3=-\frac{1}{2}(d\xi_-+\cos\omega d\xi_+),\\
        &E_1=\sigma_1-\frac{(c+H)\sin \omega}{P}(\tan \theta \sin \xi_-d\theta+\cos \xi_-\sigma_3),\\
        &E_2=\sigma_2-\frac{(c+H)\sin \omega}{P}(\tan \theta \cos \xi_-d\theta-\sin \xi_-\sigma_3),\\
    &E_3=\sigma_3.
    \end{split}
\end{equation}

The dilaton and axion are
\begin{equation}
\begin{split}
    &e^{\Phi}=\frac{g_s}{\sqrt{Q_1 Q_2}}\left(Q_1\sin^2 \phi+Q_2 \cos^2 \phi\right),\\
    &C^{(0)}=-\frac{R^2 \sinh^2 u \sin^2 \theta \sin 2\phi}{2g_s(Q_1\sin^2\phi+Q_2\cos^2\phi)}.
    \end{split}
\end{equation}

\begin{equation}
    B^{(2)}=-\cos \phi\Psi_2+\sin \phi d\phi\wedge \Psi_1,\quad g_sC^{(2)}=\sin \phi\Psi_2+\cos \phi d\phi \wedge \Psi_1,
\end{equation}
where
\begin{equation}
    \begin{split}
        &\Psi_2=\frac{2R\sinh u }{g^2\sqrt{c^2-1}}\cos \theta\left[\frac{2(1-c^2)}{P}d\theta\wedge E_3-\frac{P}{Q_1}\sin 2 \theta E_1 \wedge E_2\right],\\
        &\Psi_1=-\frac{2R\sinh u\sqrt{c^2-1}}{g^2Q_2}\cos \theta\sin 2\theta E_3.
    \end{split}
\end{equation}

The four form along the internal $S^5$ (noting that the five form field strength is given by imposing 10D self-duality) is
\begin{equation}
    C^{(4)}=\frac{8P(Q_1+Q_2)}{g_sg^4Q_1Q_2}\cos^4 \theta E_1\wedge E_2 \wedge E_3 \wedge d\phi.
\end{equation}

\subsection{Holographic Wilson loop}
Here, we analytically evaluate the holographic Wilson loop wrapping a great circle of $S^4$ using the probe fundamental string prescription of \cite{Bobev:2018hbq}, where the corresponding on-shell action was evaluated numerically.
The expression is
\begin{equation}
    S_\text{string}=\frac{1}{\ell_s^2}\int dr  e^{A} \eta^2.
\end{equation}
Using our analytic solution in Eq.~\ref{fieldssol} and introducing a UV cutoff at $u=\epsilon$, we get
\begin{equation}
\frac{4\sqrt{1-c^2}}{g^2\ell_s^2}\int_\epsilon^\infty \frac{du}{\sinh^2u} =\frac{4\sqrt{1-c^2}}{g^2\ell_s^2}\left[\coth \epsilon - 1\right].
\end{equation}

Subtracting the counterterm gives the renormalised on-shell action
\begin{equation}
    S^\text{Ren.}_\text{string}=-\frac{4\sqrt{1-c^2}}{g^2\ell_s^2},
\end{equation}
and therefore the corresponding Wilson loop expectation value
\begin{equation}
\log \langle W(C)\rangle=-S^\text{Ren.}_\text{string}=\frac{4\sqrt{1-c^2}}{g^2\ell_s^2}.
\end{equation}

\subsection{Equal-mass $\mathcal{N}=1^\ast$ on $S^4$}

A natural extension would be to consider whether the supergravity theory holographically dual to the $\mathcal{N}=1^\ast$ theory on $S^4$ \cite{Bobev:2016nua} admits a similar reduction in complexity when coordinates adapted to the scalar coset are used. The system similarly simplifies as in the $\mathcal{N}=2^\ast$ case. Although we can reduce the BPS equations to two first order coupled ODEs (with the other fields then recovered by integration or algebraic expressions), we do not find a conserved quantity that allows the full solution to be written analytically. The details of this calculation can be found in App.~\ref{app:onestar}. As shown in \cite{Bobev:2019wnf}, the large-radius limit should recover the $\lambda=1$ GPPZ solution \cite{Girardello:1999bd}.

\section{Conclusions}

In this paper, we have obtained a universal analytic solution to the BPS equations describing spherical D$p$-branes for $1\leq p\leq 5$, together with extrapolations to $p=0$ and $p=-1$. The change of variables that makes these equations integrable also greatly simplifies holographic calculations, which allowed us to evaluate the Wilson loop analytically and determine the UV asymptotics. Our second main result is an exact solution of the BPS equations describing the holographic dual of $\mathcal{N}=2^\ast$ SYM on $S^4$, together with its type IIB uplift and an analytic calculation of the holographic Wilson loop. There are many natural future questions to consider. 

For the D0-brane system holographically dual to the BMN model, we obtained a one-parameter family of analytic BPS solutions characterised by an exact conserved quantity. Imposing smooth IR boundary conditions is a singular limit in that one must send the integration constant to infinity. By expressing our solution in the variables of recent work by Santos \cite{Santos:2026rrx}, we found a perturbative description of the limiting regions observed numerically. This provides a starting point to investigate further deformations away from the BPS limit.

For the D(-1)-brane system, we similarly obtained a one-parameter family of analytic BPS solutions. Using the electrostatic formulation of holographic dual of the polarised IKKT model, we showed that no choice of integration constant leads to a regular Euclidean type IIB solution. We separately solved the one-ball electrostatic problem using an explicit conformal map from the literature, which allowed us to write multipole moments and the holographic on-shell action in terms of elliptic invariants, without numerical integration. An interesting future direction is whether any special multi-ball configurations can be solved as explicitly, despite them being generically involving hyperelliptic functions.

The D5-brane solution as well as its S-dual NS5-brane solution provide an interesting opportunity to investigate little string theory and to calculate further holographic observables. Furthermore, it would be very interesting to investigate non-supersymmetric deformations of the spherical NS5-brane solution in the style of \cite{Aharony:2019zsx}.

Having analytically solved the $\mathcal{N}=2^\ast$ on $S^4$ BPS equations by finding a conserved quantity, it would be interesting to revisit previous numerical solutions to related systems. The general method would be to work in coordinates adapted to the scalar coset manifold then try to find some invariant submanifold on which a conserved quantity can be found, then test that conserved quantity in the full system. Having applied this to equal mass $\mathcal{N}=1^\ast$, we reduced the BPS equations to two coupled first order ODEs but could not obtain an analytic solution. It would be interesting to see if this problem or the one-mass version might admit additional conserved quantities.

This paper has focused on obtaining exact supergravity solutions. The variables chosen led to particularly simple expressions for holographic Wilson loops, suggesting they might be useful in calculating further holographic observables. Our results for spherical branes and $N=2^\ast$ provide analytic tools for exploring nonconformal holography across different dimensions and theories.

\section*{Note added}
\label{sec:addednote}
Whilst preparing this manuscript, we and Fri{\dh}rik Freyr Gautason became aware that we had independently obtained the $\mathcal{N}=2^\ast$ solution presented in Section \ref{sec:nine} of this paper (see Section 5 of \cite{Gautason:2026dcl}). The results are published separately with mutual acknowledgement. 

\section*{Acknowledgements}
The author would like to thank Nikolay Bobev, Pieter Bomans, Michele Galli, Fri{\dh}rik Freyr Gautason, Michael Gutperle, Okan G\"unel, Aritra Saha and Christopher Pope for helpful and interesting discussions. The only use of AI was to correct typographical errors, speed up writing basic Mathematica code, and help find papers.

\appendix

\section{D1-branes: embedding in 3D gauged supergravity}

\label{subsec:3dembedding}

Here, we show how the spherical D1-brane theory arises as a consistent truncation of maximal 3D gauged supergravity with gauge group $SO(8)\ltimes T^{28}$, obtained by a reduction of type IIB supergravity on $S^7$ \cite{Fischbacher:2003yw,3dconsistent}. Note that we derive the Lorentzian theory, which then becomes the Euclidean theory via analytic continuation $\chi \rightarrow i \chi$ as discussed previously.

The scalars are parameterised by the coset $E_{8(8)}/SO(16)$ and transform in the $128_s$ of $SO(16)$. It branches into the relevant $SO(8)=(SO(8)\times SO(8))_\text{diag}$ gauging for type IIB supergravity as
\begin{equation}
    128=8_v\otimes 8_v +8_s \otimes 8_c=1\oplus 28\oplus 35_v\oplus 8_v\oplus 56_v \enspace .
\end{equation}
This gives three $SO(3)\times SO(5)$ singlets, one in each of $1$, $35_v$, and $56_v$.

It is convenient to decompose $E_{8(8)}$ into $SL(9)$ 
\begin{equation}
    248=80\oplus 84\oplus \overline{84},
\end{equation}
where the generators are represented by traceless $9\times 9$ matrices and antisymmetric 3-forms. In particular, we use the same conventions and normalisations as Eqs.3.98-3.100 of \cite{Berman:2019efr}, with generators $E^m_n,Z^{m_1,m_2,m_3},Z_{m_1,m_2,m_3}$ with indices running from $1$ to $9$. 

The three $SO(3)\times SO(5)$ singlets correspond to the generators
\begin{equation}
\begin{split}
    &N=Y^{cc}\equiv \frac{1}{3} \text{diag}(1,1,1,1,1,1,1,1,-8),\\
    & H\equiv \text{diag} (5,5,5,-3,-3,-3,-3,-3,0),\\
    & Z^{123}+Z_{123}.
\end{split}
\end{equation}

$N$ is the singlet under $SO(8)$. $Z^{123}+Z_{123}$ is the singlet within $56_v$ which is the volume form on $SO(3)$ so is invariant under $SO(3)\times SO(5)$. $H$ is in $35_v$.

We use the embedding tensor in Eq.56. of \cite{Fischbacher:2003yw}
\begin{equation}
    \Theta=E_{ab} \vee Z_{ab},
\end{equation}
using the symmetric tensor product $\vee$, where we identify their variables on the left with the $SL(9)$ generators on the right as
\begin{equation}
    Z_{ab}=Z_{ab9}, \quad E_{ab}=E^{a}_b-E^b_a.
\end{equation}

Using the group representative
\begin{equation}
    \mathcal{V}=\exp(\chi Z^{123})\exp\left(-\frac{\beta}{16}(N+H)+\frac{\eta}{80}(H-15N)\right),
\end{equation}
and the embedding tensor of \cite{Fischbacher:2003yw}, we substitute these into the Lagrangian in Eq.4.14/4.15 of~\cite{Eloy:2023zzh} to get
\begin{equation}
    \mathcal{L}=-\frac{1}{2}(d \beta)^2-\frac{3}{10}(d \eta)^2-\frac{1}{2}e^{2\beta}(d\chi)^2+\frac{3}{2}g^2e^{-\beta+\frac{3}{5}\eta}\left(5e^{2\beta}+10e^{\beta+\frac{\eta}{5}}+e^{\frac{2}{5}\eta}(1+\chi^2e^{2\beta})\right), \label{d1lag}
\end{equation}
which is precisely the Lagrangian for $p=1$.

By reducing to all singlets under $(SO(3)\times SO(5))$, we get a consistent bosonic truncation of maximal 3D gauged supergravity. We do not derive the supersymmetry variations here.

\section{Derivation of one-ball solution}

Here, we provide a derivation of the solution presented in Section \ref{subsec:oneballsol}. Useful identities and expansions for elliptic functions can be found in \cite{NIST:DLMF}.

\label{app:oneball}
\subsection{Elliptic invariants}
Since we want to highlight how the solutions can be written explicitly without integrals, here we give a definition of elliptic invariants we will use. For periods $\omega_1=1$ and $\omega_2=im$ and $\eta_1\equiv2 \zeta(1/2)$, if we define
\begin{equation}
    q=e^{-2\pi m},
\end{equation}
we can write
\begin{equation}
    \begin{split}
        &\eta_1(m)=\frac{\pi^2}{3}\left(1-24\sum_{n=1}^\infty \frac{nq^n}{1-q^n}\right),\\
        &g_2(m)=\frac{4\pi^4}{3}\left(1+240\sum_{n=1}^\infty \frac{n^3 q^n}{1-q^n}\right),\\
        &g_3(m)=\frac{8\pi^6}{27}\left(1-504\sum_{n=1}^\infty \frac{n^5q^n}{1-q^n}\right).
        \end{split}
\end{equation}

\subsection{Laplace equation}

In the electrostatic formulation \cite{komatsu1}, $\rho V$ satisfies the 2D Laplace equation, so locally we can write
\begin{equation}
    W=\rho V=\text{Re}(F(Z)), \quad Z=\rho+iz,
\end{equation}
for a holomorphic $F$.

We consider a ball with linear charge density $f_s(u)$ (defined in Eq.5.1. of \cite{komatsu2}) centered at $z_s$ with radius $R_s$. Including its image at $z=-z_s$ to ensure that $V(\rho,0)=0$, we get
\begin{equation}
  \rho V_{\text{ball}}=\text{Re}\left[\frac{1}{4\pi^2 }\int _{-R_s}^{R_s} du \frac{f_s(u)}{Z-u-iz_s}-\frac{1}{4\pi^2 }\int _{-R_s}^{R_s} du \frac{f_s(u)}{Z-u+iz_s}\right]\equiv \text{Re}[F_s]. \label{integralexp}
\end{equation}

\subsection{Conformal mapping}

$F_s$ is holomorphic in the complex plane minus two slits, which we denote $G$, so
\begin{equation}
    G=\mathbb{C}\backslash ([-R_s,R_s]\pm i z_s).
\end{equation}

We can then recognise this as the situation described at the beginning of page 14 of \cite{doubleslit}, with $b=z_s$, $l=2R_s$. In particular, we can then write an explicit conformal map 
\begin{equation}
\begin{split}
&\Pi\rightarrow G: v\rightarrow Z;\\
    &Z(v)=-\frac{z_s}{\pi}(\zeta(v)-\eta_1v),\\
    &\text{where }\Pi=\{-\frac{1}{2}<\text{Re}(v)<\frac{1}{2},-\frac{m}{2}<\text{Im}(v)<\frac{m}{2}\},
\end{split}
\end{equation}
for $\zeta$ the Weierstrass zeta function with periods $\omega_1=1$ and $\omega_2=im$, and $\eta_1\equiv 2\zeta(1/2)$.

The two horizontal boundaries of $\Pi$ are mapped to the two slits and the vertical boundaries of $\Pi$ are identified so that $Z(v+1)=Z(v)$.

Physical infinity in $G$ corresponds to $v=0$ since
\begin{equation}
    Z(v)=-\frac{z_s}{\pi }\frac{1}{v}+O(v).
\end{equation}

The critical point $v=v_c$ of the map $v\rightarrow Z(v)$ is given by
\begin{equation}
    \wp(v_c)=-\eta_1,
\end{equation}
for the Weierstrass elliptic function $\wp$ which then defines $R_s$ through
\begin{equation}
    R_s=\frac{z_s}{\pi}|\text{Re}[\zeta(v_c)-\eta_1v_c]|.
\end{equation}

\subsection{Including a background}

We consider the background that gives the correct asymptotics for the polarised IKKT model
\begin{equation}
    V_{bg}=-\eta z \mu^3+\frac{\mu^5}{2^7}(z\rho^2-z^3).
\end{equation}
We then write it as
\begin{equation}
    \rho V_{bg}=\text{Re}\left[\frac{\eta \mu^3}{2} iZ^2-\frac{\mu^5}{2^{9}}iZ^4\right]\equiv \text{Re}[{F_{bg}}].
\end{equation}

\subsection{Boundary conditions}

However, $V_{bg}$ does not obey $V(\rho,z_s)=V_s$ for some constant $V_s$ for $0<\rho < R_s$. Therefore, we want to write something which obeys this condition whilst giving the same asymptotics (now mapped to $v\rightarrow 0$). To do this, we swap $Z^4$ for a linear combination of $Z_v^2$ and $Z^2$ with coefficients tuned to give the same asymptotic expansion whilst giving the correct boundary conditions.

Using the expansion as $v\rightarrow 0$,
\begin{equation}
    Z=-\frac{z_s}{v\pi}+\frac{z_s}{\pi}\eta_1v+\frac{z_s}{60\pi}g_2v^3+\mathcal{O}(v^5),
\end{equation}
choosing
\begin{equation}
    F(Z(v))=\left(\frac{\eta\mu^3}{2}+\frac{3\mu^5z_s^2\eta_1}{2^8\pi^2}\right)iZ(v)^2-\frac{z_s^2 \mu^5}{2^9\pi^2}iZ_v(v)^2,
\end{equation}
gives the correct asymptotics matching the desired background.

On the horizontal boundaries of $\Pi$, $v=x\pm i\frac{m}{2}$. Consider one of them $Z(v)=\rho(x) \pm iz_s$, so 
\begin{equation}
    Z_v=\frac{dZ}{dx}\frac{dx}{dv}=\frac{dZ}{dx}=\frac{d\rho}{dx}\in \mathbb{R},
\end{equation}
which implies
\begin{equation}
    \text{Re}[iZ_v^2]=0, \quad \text{Re}[iZ^2]=-2\rho z_s,
\end{equation}
so $Z_v$ does not contribute to $V_s$ and $V_s$ can be written explicitly as
\begin{equation}
    V_s=-\eta\mu^3z_s-\frac{3\mu^5 z_s^3\eta_1}{2^7\pi^2}.
\end{equation}
A similar argument applies to the other horizontal boundary.

\subsection{Multipole expansion}
We consider $F_s$ to correspond to the potential with the background subtracted, so
\begin{equation}
    F_s=\frac{3\mu^5z_s^2\eta_1}{2^8\pi^2}iZ^2+\frac{\mu^5}{2^{9}}iZ^4-\frac{z_s^2 \mu^5}{2^9\pi^2}iZ_v^2.
\end{equation}
In the language of \cite{komatsu1}, we want to calculate the charge $Q_s$ and multipoles $q_n$ using
\begin{equation}
    F=\frac{\eta \mu^3}{2} iZ^2-\frac{\mu^5}{2^{9}}iZ^4+\frac{iz_sQ_s}{2\pi^2Z^2}-\frac{iq_3}{4\pi^2Z^4}+\frac{iq_5}{4\pi^2Z^6}+\mathcal{O}(Z^{-8}).
\end{equation}

Using the expansion of $Z(v)$ as $v\rightarrow 0$
\begin{equation}
    Z(v)=\left(-\frac{z_s}{v\pi}+\frac{z_s}{\pi}\eta_1v+\frac{z_s}{60\pi}g_2v^3+\frac{z_s}{140\pi}g_3v^5+\frac{z_s}{8400\pi}g_2^2v^7\right)+\mathcal{O}(v^9),\label{zexp}
\end{equation}
we then get
\begin{equation}
\begin{split}
    F_s\rightarrow &-\frac{(42\eta_1^2+g_2)\mu^5z_s^4}{3072\pi^4}i+\frac{i(20\eta_1^3-\eta_1g_2-g_3)\mu^5z_s^4}{5120\pi^4}v^2\\
    &+\frac{i(336\eta_1^4-g_2^2-24\eta_1g_3)\mu^5z_s^4}{172032\pi^4}v^4+\mathcal{O}(v^6).
\end{split}
\end{equation}

So we can match coefficients to get
\begin{equation}
Q_s=\frac{z_s^5\mu^5(20\eta_1^3-\eta_1g_2-g_3)}{2560\pi^4}.
\end{equation}
\begin{equation}
    q_3=\frac{(5040\eta_1^4-336\eta_1^2g_2+5g_2^2-216\eta_1g_3)\mu^5z_s^8}{215040\pi^6}.
\end{equation}

Note that up to arbitrary order, one can determine the higher order multipoles in terms of elliptic invariants. A similar calculation gives
\begin{equation}
q_5=\frac{\eta_1(504\eta_1^4-42\eta_1^2g_2+g_2^2-18\eta_1g_3)\mu^5z_s^{10}}{10752\pi^8}.
\end{equation}

\subsection{On-shell action}
\label{subsec:onshell}
These quantities allow us to calculate the on-shell action of the matrix model as in Eq.5.49. of \cite{komatsu2} (noting that it was only defined there for $\eta=0$)
\begin{equation}
    S_\text{on-shell}=\frac{2^8}{3\pi^4g_s^2\alpha'^4\mu^{10}}QV-\frac{1}{3g_s^2\pi^4\alpha'^4\mu^5}q_3.
\end{equation}

Using our expressions, we get
\begin{equation}
\begin{split}
    S_\text{on-shell}=-\frac{z_s^8}{129024\pi^{10}g_s^2\alpha'^4}&\left(3024\eta_1^4-168\eta_1^2g_2-144\eta_1g_3+g_2^2\right).
\end{split}
\end{equation}

We also have
\begin{equation}
    \xi=\frac{3^5}{2^{20}\pi^2}\frac{Q_s}{z_s^5 \mu^5}=\frac{3^5}{2^{20}\pi^6 2560}(20\eta_1^3-\eta_1g_2-g_3).
\end{equation}

Then
\begin{equation}
    H(\xi)=-\frac{\lambda^\frac{2}{3}}{N^2}S_\text{on-shell},
\end{equation}
can be written parametrically with $\xi$ and $H(\xi)$ both written in terms of elliptic invariants.

\subsection{Charge density at tip}

At the tip of the ball, $Z(v_c)=R_s+iz_s$ and $Z_v(v_c)=0$.

Using our potential, we have
\begin{equation}
\frac{dF}{dZ}=2\left(\frac{\eta\mu^3}{2}+\frac{3\mu^5z_s^2\eta_1}{2^8\pi^2}\right)iZ_c-2\frac{z_s^2 \mu^5}{2^9\pi^2}iZ_{vv}(v_c).
\end{equation}
Since $Z$ and $Z_{vv}$ are both analytic at $v=v_c$, we have no jump as $z$ tends to $z_s$ from above and below as $\rho \rightarrow R_s$.
Since
\begin{equation}
    V_z=-\left(\frac{1}{\rho}\right)\text{Im}\left(\frac{dF}{dZ}\right),
\end{equation}
the charge density vanishes at the tip as $\rho \rightarrow R_s$, as required by \cite{komatsu1}.

\section{Equal-mass $\mathcal{N}=1^\ast$ on $S^4$}
\label{app:onestar}
With the metric
\begin{equation}
    ds^2=dr^2+e^{2A(r)}d\Omega_4^2,
\end{equation}
the Lagrangian is
\begin{equation}
\begin{split}
    \mathcal{L}=&-\frac{R}{4}-\frac{1}{8}((\partial Y^0)^2-(\partial Y^1)^2+(\partial Y^2)^2)-\frac{3}{8}((\partial Z^0)^2-(\partial Z^1)^2+(\partial Z^2)^2)\\
    &-\frac{3}{4}(3-2(Z^{2})^2+(1+(Z^{2})^2)(Y^0Z^0+Y^1Z^1)+Y^2({Z^{2}})^3).
    \end{split}
\end{equation}
We then use similar embedding coordinates as we used for the $\mathcal{N}=2^\ast$ on $S^4$ case.
\begin{equation}
\begin{split}
&Y^0=\frac{1+z_1\tilde{z_1}}{1-z_1\tilde{z_1}}, \quad Y^1=\frac{z_1+\tilde{z_1}}{1-z_1\tilde{z_1}}, \quad Y^2=\frac{z_1-\tilde{z_1}}{1-z_1\tilde{z_1}},\\
&Z^0=\frac{1+z_2\tilde{z_2}}{1-z_2\tilde{z_2}}, \quad Z^1=\frac{z_2+\tilde{z_2}}{1-z_2\tilde{z_2}}, \quad Z^2=\frac{z_2-\tilde{z_2}}{1-z_2\tilde{z_2}},\\
&(Y^{0})^2-(Y^1)^2+(Y^2)^2=1,\\
&(Z^{0})^2-(Z^1)^2+(Z^2)^2=1.
\end{split}
\end{equation}

Define for convenience the quantity
\begin{equation}
    \Sigma=Y^0Z^0+Y^1Z^1+Y^2Z^2,
\end{equation}
then the BPS equations in embedding coordinates are
\begin{equation}
\begin{split}
&\frac{dY^0}{dr}=\frac{3}{1+\Sigma}(A'(Z^0-\Sigma Y^0)+e^{-A}(Z^2Y^1+Y^2Z^1)),\\
&\frac{dY^1}{dr}=\frac{3}{1+\Sigma}(A'(-Z^1-\Sigma Y^1)+e^{-A}(Z^2Y^0-Y^2Z^0)),\\
&\frac{dY^2}{dr}=\frac{3}{1+\Sigma}(A'(Z^2-\Sigma Y^2)-e^{-A}(Y^0Z^1+Y^1Z^0)),\\
&\frac{dZ^0}{dr}=\frac{A'(Y^0-\Sigma Z^0)+e^{-A}(Y^2Z^1+Z^2Y^1)}{1+\Sigma}+\frac{2Z^2}{1-(Z^2)^2}\left(A'Z^2Z^0-e^{-A}Z^1\right),\\
&\frac{dZ^1}{dr}=\frac{A'(-Y^1-\Sigma Z^1)+e^{-A}(Y^2Z^0-Z^2Y^0)}{1+\Sigma}+\frac{2Z^2}{1-(Z^2)^2}\left(A'Z^2Z^1-e^{-A}Z^0\right),\\
&\frac{dZ^2}{dr}=\frac{A'(Y^2-\Sigma Z^2)-e^{-A}(Y^0Z^1+Y^1Z^0)}{1+\Sigma}-2A'Z^2,\\
&\left(\frac{dA}{dr}\right)^2=e^{-2A}+\frac12\left(1-(Z^2)^2\right)\left(1+Y^0Z^0+Y^1Z^1+Y^2Z^2\right).
\end{split}
\end{equation}

Experimenting with linear combinations of these variables gives
\begin{equation}
\frac{d}{dr}(Y^2-3Z^2)
=-3A'(Y^2-3Z^2).
\end{equation}
This gives the conserved quantity
\begin{equation}
e^{3A}(Y^2-3Z^2)=\kappa.
\end{equation}

For a regular solution, the $S^4$ shrinks smoothly in the IR, so $e^A\rightarrow 0$ while the scalar fields remain finite. This requires $\kappa=0$, which implies
\begin{equation}
Y^2=3Z^2.
\end{equation}

We can therefore make the parameterisation
\begin{equation}
\begin{split}
    &Y^0=\sqrt{1-9w^2}\cosh u, \quad Y^1=\sqrt{1-9w^2}\sinh u, \quad Y^2=3w,\\
    & Z^0 = \sqrt{1-w^2}\cosh v, \quad Z^1=\sqrt{1-w^2} \sinh v, \quad Z^2=w. \label{hypers}
\end{split}
\end{equation}

Define for convenience 
\begin{equation}
    N(w)=\sqrt{(1-w^2)(1-9w^2)}
\end{equation}

The BPS equations become

\begin{equation}
    \begin{split}
        &\frac{du}{dr}=\frac{3[e^{-A}w(1-9w^2-3N\cosh(u+v))-A'N\sinh(u+v)]}{(1-9w^2)[1+3w^2+N\cosh(u+v)]},\\
        &\frac{dv}{dr}=\frac{[e^{-A}w(1-9w^2-3N\cosh(u+v))-A'N\sinh(u+v)]}{(1-w^2)[1+3w^2+N\cosh(u+v)]},\\
        &\frac{dw}{dr}=\frac{A'w(1-9w^2-3N\cosh(u+v))-e^{-A}N\sinh(u+v)}{1+3w^2+N\cosh(u+v)},\\
        &A'^2=e^{-2A}+\frac{1-w^2}{2}\left[1+3w^2+N\cosh(u+v)\right].
    \end{split}
\end{equation}

Since only $(u+v)$ appears on the right hand side of these equations and 
\begin{equation}
    (1-9w^2)\frac{du}{dr}=3(1-w^2)\frac{dv}{dr},
\end{equation}
we define
\begin{equation}
    \sigma=u+v,
\end{equation}
then
\begin{equation}
\frac{du}{dr}=\frac{3(1-w^2)}{4(1-3w^2)}\sigma', \quad \quad \quad \frac{dv}{dr}=\frac{1-9w^2}{4(1-3w^2)}\sigma'. \label{uvw}
\end{equation}

Doing this means that, if we solve for $\sigma$, then $u$ and $v$ can be given in terms of an integral with respect to $w$ using Eqs.~\ref{uvw}.

\begin{equation}
    \begin{split}
        &\frac{d\sigma}{dr}=\frac{4(1-3w^2)[e^{-A}w(1-9w^2-3N\cosh\sigma)-A'N\sinh\sigma]}{(1-w^2)(1-9w^2)[1+3w^2+N\cosh\sigma]},\\
        &\frac{dw}{dr}=\frac{A'w(1-9w^2-3N\cosh\sigma)-e^{-A}N\sinh\sigma}{1+3w^2+N\cosh \sigma},\\
        &A'^2=e^{-2A}+\frac{1-w^2}{2}\left[1+3w^2+N\cosh \sigma\right].
    \end{split}
\end{equation}

Let
\begin{equation}
    J=\frac{1-w^2}{2}\left[1+3w^2+\sqrt{(1-9w^2)(1-w^2)}\cosh \sigma\right].
\end{equation}

We can choose to use $A$ itself as a radial coordinate to get two coupled first order ODEs
\begin{equation}
    \begin{split}
        &\frac{dw}{dA}=\frac{w}{J}(2-2w^2-3J)\mp\frac{\sqrt{J^2-J(1-w^2)(1+3w^2)+4w^2(1-w^2)^2}}{J\sqrt{1+e^{2A}J}},\\
       & \frac{dJ}{dA}=4(1-3w^2)\left(1-\frac{J}{1-w^2}\right),
    \end{split}
\end{equation}
with the $\mp$ sign simply reflecting a choice of sign of $\sinh \sigma$.

If these two ODEs are solved, $\sigma$ follows algebraically, $u$ and $v$ follow by integration in Eqs.~\ref{uvw}, then the full BPS ODEs will be solved. However, although these equations are simpler than we started, they still do not seem to be tractable.

 \bibliographystyle{utphys}
\bibliography{biblip}

@article{spherical1,
    author = "Bobev, Nikolay and Bomans, Pieter and Gautason, Fri{\dh}rik Freyr",
    title = "{Spherical Branes}",
    eprint = "1805.05338",
    archivePrefix = "arXiv",
    primaryClass = "hep-th",
    doi = "10.1007/JHEP08(2018)029",
    journal = "JHEP",
    volume = "08",
    pages = "029",
    year = "2018"
}

@inproceedings{henning,
    author = "Ciceri, Franz and Samtleben, Henning",
    title = "{The matrix edge of holography}",
    eprint = "2604.00355",
    archivePrefix = "arXiv",
    primaryClass = "hep-th",
    month = "4",
    year = "2026"
}

@article{henning2,
    author = "Ciceri, Franz and Samtleben, Henning",
    title = "{Supergravity dual for Ishibashi-Kawai-Kitazawa-Tsuchiya holography}",
    eprint = "2511.23111",
    archivePrefix = "arXiv",
    primaryClass = "hep-th",
    doi = "10.1103/gmhl-mmg5",
    journal = "Phys. Rev. D",
    volume = "113",
    number = "4",
    pages = "046001",
    year = "2026"
}

@article{spherical2,
  title="{Spherical branes and the BMN matrix quantum mechanics}",
  author={Bobev, Nikolay and Bomans, Pieter and Gautason, Fri{\dh}rik Freyr},
  journal={Journal of High Energy Physics},
  volume={2025},
  number={1},
  archivePrefix={arXiv},
  primaryClass={hep-th},
  eprint={2410.21376},
  pages={170},
  year={2025},
  publisher={Springer}
}

@article{lst1,
  title={A brief review of 'little string theories'},
  author={Aharony, Ofer},
  journal={Classical and Quantum Gravity},
  volume={17},
  number={5},
  pages={929--938},
  year={2000}
}

@article{twostar1,
    author = "Bobev, Nikolay and Elvang, Henriette and Freedman, Daniel Z. and Pufu, Silviu S.",
    title = "{Holography for $N = 2^*$ on $S^4$}",
    eprint = "1311.1508",
    archivePrefix = "arXiv",
    primaryClass = "hep-th",
    reportNumber = "MCTP-13-36, MIT-CTP-4513",
    doi = "10.1007/JHEP07(2014)001",
    journal = "JHEP",
    volume = "07",
    pages = "001",
    year = "2014"
}

@article{Aharony:2019zsx,
    author = "Aharony, Ofer and Evtikhiev, Mikhail and Feldman, Andrey",
    title = "{Little String Theories on Curved Manifolds}",
    eprint = "1908.02642",
    archivePrefix = "arXiv",
    primaryClass = "hep-th",
    doi = "10.1007/JHEP10(2019)180",
    journal = "JHEP",
    volume = "10",
    pages = "180",
    year = "2019"
}

@misc{doubleslit,
      title={Conformal module of the exterior of two rectilinear slits}, 
      author={D. Dautova and S. Nasyrov and M. Vuorinen},
      year={2019},
      eprint={1908.02459},
      archivePrefix={arXiv},
      primaryClass={math.CV},
      url={https://arxiv.org/abs/1908.02459}, 
}

@article{komatsu1,
    author = "Komatsu, Shota and Martina, Adrien and Penedones, Jo{\~a}o and Vuignier, Antoine and Zhao, Xiang",
    title = "{Einstein gravity from a matrix integral. Part I}",
    eprint = "2410.18173",
    archivePrefix = "arXiv",
    primaryClass = "hep-th",
    doi = "10.1007/JHEP12(2025)029",
    journal = "JHEP",
    volume = "12",
    pages = "029",
    year = "2025"
}

@article{komatsu2,
    author = "Komatsu, Shota and Martina, Adrien and Penedones, Joao and Vuignier, Antoine and Zhao, Xiang",
    title = "{Einstein gravity from a matrix integral. Part II}",
    eprint = "2411.18678",
    archivePrefix = "arXiv",
    primaryClass = "hep-th",
    doi = "10.1007/JHEP12(2025)030",
    journal = "JHEP",
    volume = "12",
    pages = "030",
    year = "2025"
}

@article{ikkt,
   title={A large-N reduced model as superstring},
   volume={498},
   ISSN={0550-3213},
   url={http://dx.doi.org/10.1016/S0550-3213(97)00290-3},
   DOI={10.1016/s0550-3213(97)00290-3},
   number={1-2},
   journal={Nuclear Physics B},
   publisher={Elsevier BV},
   author={Ishibashi, Nobuyuki and Kawai, Hikaru and Kitazawa, Yoshihisa and Tsuchiya, Asato},
   year={1997},
   month=Aug, pages={467–491} }

@article{ikktdual1,
    author = "Ooguri, Hirosi and Skenderis, Kostas",
    title = "{On the field theory limit of D instantons}",
    eprint = "hep-th/9810128",
    archivePrefix = "arXiv",
    reportNumber = "UCB-PTH-98-49A, LBNL-42387, SPIN-1998-01, LBL-42387",
    doi = "10.1088/1126-6708/1998/11/013",
    journal = "JHEP",
    volume = "11",
    pages = "013",
    year = "1998"
}

@article{ikktdual2,
    author = "Gibbons, Gary W. and Green, Michael B. and Perry, Malcolm J.",
    title = "{Instantons and seven-branes in type IIB superstring theory}",
    eprint = "hep-th/9511080",
    archivePrefix = "arXiv",
    reportNumber = "DAMTP-R-95-56",
    doi = "10.1016/0370-2693(95)01565-5",
    journal = "Phys. Lett. B",
    volume = "370",
    pages = "37--44",
    year = "1996"
}

@article{ikktdual3,
    author = "Bergshoeff, E. and Behrndt, K.",
    title = "{D - instantons and asymptotic geometries}",
    eprint = "hep-th/9803090",
    archivePrefix = "arXiv",
    reportNumber = "UG-2-98, HUB-EP-14-98",
    doi = "10.1088/0264-9381/15/7/002",
    journal = "Class. Quant. Grav.",
    volume = "15",
    pages = "1801--1813",
    year = "1998"
}

@article{polarisedikkt,
    author = "Hartnoll, Sean A. and Liu, Jun",
    title = "{The polarised IKKT matrix model}",
    eprint = "2409.18706",
    archivePrefix = "arXiv",
    primaryClass = "hep-th",
    doi = "10.1007/JHEP03(2025)060",
    journal = "JHEP",
    volume = "03",
    pages = "060",
    year = "2025"
}

@article{gutperle2,
    author = "D'Hoker, Eric and Gutperle, Michael and Uhlemann, Christoph F.",
    title = "{Type IIB supergravity solutions for polarized IKKT}",
    eprint = "2609.35963",
    archivePrefix = "arXiv",
    primaryClass = "hep-th",
    month = "9",
    year = "2026"
}

@article{3dconsistent,
    author = "Galli, Michele and Malek, Emanuel",
    title = "{Consistent truncations to 3-dimensional supergravity}",
    eprint = "2206.03507",
    archivePrefix = "arXiv",
    primaryClass = "hep-th",
    reportNumber = "HU-EP-22/23",
    doi = "10.1007/JHEP09(2022)014",
    journal = "JHEP",
    volume = "09",
    pages = "014",
    year = "2022"
}

@article{Santos:2026rrx,
    author = "Santos, Jorge E.",
    title = "{Blackened BPS: the end of the thermal BMN branch}",
    eprint = "2610.02305",
    archivePrefix = "arXiv",
    primaryClass = "hep-th",
    month = "10",
    year = "2026"
}

@article{Pestun:2007rz,
    author = "Pestun, Vasily",
    title = "{Localization of gauge theory on a four-sphere and supersymmetric Wilson loops}",
    eprint = "0712.2824",
    archivePrefix = "arXiv",
    primaryClass = "hep-th",
    reportNumber = "ITEP-TH-41-07, PUTP-2248",
    doi = "10.1007/s00220-012-1485-0",
    journal = "Commun. Math. Phys.",
    volume = "313",
    pages = "71--129",
    year = "2012"
}

@article{Bobev:2018hbq,
    author = "Bobev, Nikolay and Gautason, Fri{\dh}rik Freyr and Van Muiden, Jesse",
    title = "{Precision Holography for $\mathcal{N}=2^{*}$ on $S^4$ from type IIB Supergravity}",
    eprint = "1802.09539",
    archivePrefix = "arXiv",
    primaryClass = "hep-th",
    doi = "10.1007/JHEP04(2018)148",
    journal = "JHEP",
    volume = "04",
    pages = "148",
    year = "2018"
}

@article{Bobev:2016nua,
    author = "Bobev, Nikolay and Elvang, Henriette and Kol, Uri and Olson, Timothy and Pufu, Silviu S.",
    title = "{Holography for $ \mathcal{N} $ = 1$^{\ast}$ on S$^{4}$}",
    eprint = "1605.00656",
    archivePrefix = "arXiv",
    primaryClass = "hep-th",
    reportNumber = "PUPT-2502, MCTP-16-12",
    doi = "10.1007/JHEP10(2016)095",
    journal = "JHEP",
    volume = "10",
    pages = "095",
    year = "2016"
}

@article{Lin:2005nh,
    author = "Lin, Hai and Maldacena, Juan Martin",
    title = "{Fivebranes from gauge theory}",
    eprint = "hep-th/0509235",
    archivePrefix = "arXiv",
    reportNumber = "PUPT-2172",
    doi = "10.1103/PhysRevD.74.084014",
    journal = "Phys. Rev. D",
    volume = "74",
    pages = "084014",
    year = "2006"
}

@article{Blau:2000xg,
    author = "Blau, Matthias",
    title = "{Killing spinors and SYM on curved spaces}",
    eprint = "hep-th/0005098",
    archivePrefix = "arXiv",
    doi = "10.1088/1126-6708/2000/11/023",
    journal = "JHEP",
    volume = "11",
    pages = "023",
    year = "2000"
}

@article{Kanitscheider:2008kd,
    author = "Kanitscheider, Ingmar and Skenderis, Kostas and Taylor, Marika",
    title = "{Precision holography for non-conformal branes}",
    eprint = "0807.3324",
    archivePrefix = "arXiv",
    primaryClass = "hep-th",
    reportNumber = "ITFA-2008-25",
    doi = "10.1088/1126-6708/2008/09/094",
    journal = "JHEP",
    volume = "09",
    pages = "094",
    year = "2008"
}

@article{Itzhaki:1998dd,
    author = "Itzhaki, Nissan and Maldacena, Juan Martin and Sonnenschein, Jacob and Yankielowicz, Shimon",
    title = "{Supergravity and the large N limit of theories with sixteen supercharges}",
    eprint = "hep-th/9802042",
    archivePrefix = "arXiv",
    reportNumber = "TAUP-2474-98, HUTP-98-A003",
    doi = "10.1103/PhysRevD.58.046004",
    journal = "Phys. Rev. D",
    volume = "58",
    pages = "046004",
    year = "1998"
}

@article{Minahan:2015jta,
    author = "Minahan, Joseph A. and Zabzine, Maxim",
    title = "{Gauge theories with 16 supersymmetries on spheres}",
    eprint = "1502.07154",
    archivePrefix = "arXiv",
    primaryClass = "hep-th",
    doi = "10.1007/JHEP03(2015)155",
    journal = "JHEP",
    volume = "03",
    pages = "155",
    year = "2015"
}

@article{Bobev:2019bvq,
    author = "Bobev, Nikolay and Bomans, Pieter and Gautason, Fri{\dh}rik Freyr and Minahan, Joseph A. and Nedelin, Anton",
    title = "{Supersymmetric Yang-Mills, Spherical Branes, and Precision Holography}",
    eprint = "1910.08555",
    archivePrefix = "arXiv",
    primaryClass = "hep-th",
    doi = "10.1007/JHEP03(2020)047",
    journal = "JHEP",
    volume = "03",
    pages = "047",
    year = "2020"
}

@article{Maldacena:1997re,
    author = "Maldacena, Juan Martin",
    title = "{The Large $N$ limit of superconformal field theories and supergravity}",
    eprint = "hep-th/9711200",
    archivePrefix = "arXiv",
    reportNumber = "HUTP-97-A097, HUTP-98-A097",
    doi = "10.4310/ATMP.1998.v2.n2.a1",
    journal = "Adv. Theor. Math. Phys.",
    volume = "2",
    pages = "231--252",
    year = "1998"
}

@article{Berenstein:2002jq,
    author = "Berenstein, David Eliecer and Maldacena, Juan Martin and Nastase, Horatiu Stefan",
    title = "{Strings in flat space and pp waves from N=4 superYang-Mills}",
    eprint = "hep-th/0202021",
    archivePrefix = "arXiv",
    doi = "10.1088/1126-6708/2002/04/013",
    journal = "JHEP",
    volume = "04",
    pages = "013",
    year = "2002"
}

@article{Kanitscheider:2009as,
    author = "Kanitscheider, Ingmar and Skenderis, Kostas",
    title = "{Universal hydrodynamics of non-conformal branes}",
    eprint = "0901.1487",
    archivePrefix = "arXiv",
    primaryClass = "hep-th",
    reportNumber = "ITFA-2009-01, NSF-KITP-2009-02",
    doi = "10.1088/1126-6708/2009/04/062",
    journal = "JHEP",
    volume = "04",
    pages = "062",
    year = "2009"
}

@article{Wiseman:2008qa,
    author = "Wiseman, Toby and Withers, Benjamin",
    title = "{Holographic renormalization for coincident Dp-branes}",
    eprint = "0807.0755",
    archivePrefix = "arXiv",
    primaryClass = "hep-th",
    doi = "10.1088/1126-6708/2008/10/037",
    journal = "JHEP",
    volume = "10",
    pages = "037",
    year = "2008"
}

@article{Najafian:2026wld,
    author = "Najafian, Arsam and Van Raamsdonk, Mark and Vilar L{\'o}pez, Alejandro",
    title = "{Compact holography via polarized IKKT}",
    eprint = "2611.03807",
    archivePrefix = "arXiv",
    primaryClass = "hep-th",
    month = "11",
    year = "2026"
}

@article{Boonstra_1999,
   title={The domain-wall/QFT correspondence},
   volume={1999},
   ISSN={1029-8479},
   url={http://dx.doi.org/10.1088/1126-6708/1999/01/003},
   DOI={10.1088/1126-6708/1999/01/003},
   number={01},
   journal={Journal of High Energy Physics},
   publisher={Springer Science and Business Media LLC},
   author={Boonstra, Harm J and Skenderis, Kostas and Townsend, Paul K},
   year={1999},
   month=Jan, pages={003–003} }

@article{Cvetic_1999,
   title={Domain-wall supergravities from sphere reduction},
   volume={560},
   ISSN={0550-3213},
   url={http://dx.doi.org/10.1016/S0550-3213(99)00445-9},
   DOI={10.1016/s0550-3213(99)00445-9},
   number={1-3},
   journal={Nuclear Physics B},
   publisher={Elsevier BV},
   author={Cvetič, M. and Liu, James T. and Lü, H. and Pope, C.N.},
   year={1999},
   month=Nov, pages={230–256} }

@article{Gautason:2026dcl,
    author = "Gautason, Fridrik Freyr",
    title = "{New analytic Janus and $S^4$ solutions for $\mathcal{N}=2^*$}",
    eprint = "2610.17238",
    archivePrefix = "arXiv",
    primaryClass = "hep-th",
    month = "10",
    year = "2026"
}

@article{Pilch:2000ue,
    author = "Pilch, Krzysztof and Warner, Nicholas P.",
    title = "{N=2 supersymmetric RG flows and the IIB dilaton}",
    eprint = "hep-th/0004063",
    archivePrefix = "arXiv",
    reportNumber = "CITUSC-00-018, USC-00-02",
    doi = "10.1016/S0550-3213(00)00656-8",
    journal = "Nucl. Phys. B",
    volume = "594",
    pages = "209--228",
    year = "2001"
}

@article{Kim:2019feb,
    author = "Kim, Nakwoo",
    title = "{Solving Mass-deformed Holography Perturbatively}",
    eprint = "1902.00418",
    archivePrefix = "arXiv",
    primaryClass = "hep-th",
    doi = "10.1007/JHEP04(2019)053",
    journal = "JHEP",
    volume = "04",
    pages = "053",
    year = "2019"
}

@article{Eloy:2023zzh,
    author = "Eloy, Camille and Galli, Michele and Malek, Emanuel",
    title = "{Adding fluxes to consistent truncations: IIB supergravity on AdS$_{3}${\texttimes} S$^{3}${\texttimes} S$^{3}${\texttimes} S$^{1}$}",
    eprint = "2306.12487",
    archivePrefix = "arXiv",
    primaryClass = "hep-th",
    reportNumber = "HU-EP-23/18",
    doi = "10.1007/JHEP11(2023)049",
    journal = "JHEP",
    volume = "11",
    pages = "049",
    year = "2023"
}

@article{Fischbacher:2003yw,
    author = "Fischbacher, T. and Nicolai, H. and Samtleben, H.",
    title = "{Nonsemisimple and complex gaugings of N=16 supergravity}",
    eprint = "hep-th/0306276",
    archivePrefix = "arXiv",
    reportNumber = "AEI-2003-050, ITP-UU-03-33, SPIN-03-21",
    doi = "10.1007/s00220-004-1081-z",
    journal = "Commun. Math. Phys.",
    volume = "249",
    pages = "475--496",
    year = "2004"
}

@article{Berman:2019efr,
    author = "Berman, David S. and Otsuki, Ray",
    title = "{Reductions of Exceptional Field Theories}",
    eprint = "1911.06150",
    archivePrefix = "arXiv",
    primaryClass = "hep-th",
    reportNumber = "QMUL-PH-19-24",
    doi = "10.1007/JHEP03(2020)066",
    journal = "JHEP",
    volume = "03",
    pages = "066",
    year = "2020"
}

@article{Girardello:1999bd,
    author = "Girardello, L. and Petrini, M. and Porrati, M. and Zaffaroni, A.",
    title = "{The Supergravity dual of N=1 superYang-Mills theory}",
    eprint = "hep-th/9909047",
    archivePrefix = "arXiv",
    reportNumber = "BICOCCA-FT-99-27, NYU-TH-99-09-02",
    doi = "10.1016/S0550-3213(99)00764-6",
    journal = "Nucl. Phys. B",
    volume = "569",
    pages = "451--469",
    year = "2000"
}

@book{Koppenfels1959,
  author    = {von Koppenfels, Werner
               and Stallmann, Friedemann},
  title     = {Praxis der konformen Abbildung},
  publisher = {Springer},
  address   = {Berlin},
  year      = {1959}
}

@article{Astesiano:2024sgi,
    author = "Astesiano, Davide and Bomans, Pieter and Gautason, Fri{\dh}rik Freyr and Puletti, Valentina Giangreco M. and Nix, Alexia",
    title = "{Wilson loops and spherical branes}",
    eprint = "2407.07830",
    archivePrefix = "arXiv",
    primaryClass = "hep-th",
    doi = "10.21468/SciPostPhys.18.2.050",
    journal = "SciPost Phys.",
    volume = "18",
    number = "2",
    pages = "050",
    year = "2025"
}

@article{Bobev:2019wnf,
    author = "Bobev, Nikolay and Gautason, Fri{\dh}rik Freyr and Niehoff, Benjamin E. and van Muiden, Jesse",
    title = "{A holographic kaleidoscope for $ \mathcal{N} $ = 1*}",
    eprint = "1906.09270",
    archivePrefix = "arXiv",
    primaryClass = "hep-th",
    doi = "10.1007/JHEP10(2019)185",
    journal = "JHEP",
    volume = "10",
    pages = "185",
    year = "2019"
}

@article{Hull:1998vg,
    author = "Hull, C. M.",
    title = "{Timelike T duality, de Sitter space, large N gauge theories and topological field theory}",
    eprint = "hep-th/9806146",
    archivePrefix = "arXiv",
    reportNumber = "QMW-PH-98-28",
    doi = "10.1088/1126-6708/1998/07/021",
    journal = "JHEP",
    volume = "07",
    pages = "021",
    year = "1998"
}

@article{Hull:1998fh,
    author = "Hull, Christopher M. and Khuri, Ramzi R.",
    title = "{Branes, times and dualities}",
    eprint = "hep-th/9808069",
    archivePrefix = "arXiv",
    reportNumber = "QMW-PH-98-37, CAMS-98-04",
    doi = "10.1016/S0550-3213(98)00691-9",
    journal = "Nucl. Phys. B",
    volume = "536",
    pages = "219--244",
    year = "1998"
}

@article{Hull:1998ym,
    author = "Hull, C. M.",
    title = "{Duality and the signature of space-time}",
    eprint = "hep-th/9807127",
    archivePrefix = "arXiv",
    reportNumber = "QMW-PH-98-30",
    doi = "10.1088/1126-6708/1998/11/017",
    journal = "JHEP",
    volume = "11",
    pages = "017",
    year = "1998"
}

@misc{NIST:DLMF,
         key = "{\relax DLMF}",
       title = "{\it NIST Digital Library of Mathematical Functions}",
howpublished = "\url{https://dlmf.nist.gov/}, Release 1.2.8 of 2026-09-15",
         url = "https://dlmf.nist.gov/",
        note = "F.~W.~J. Olver, A.~B. {Olde Daalhuis}, D.~W. Lozier, B.~I. Schneider,
                R.~F. Boisvert, C.~W. Clark, B.~R. Miller, B.~V. Saunders,
                H.~S. Cohl, and M.~A. McClain, eds."}

\end{document}